\documentclass[prb,superscriptaddress,showpacs,papersize=a4paper,twocolumn]{revtex4-1}
\usepackage[pdftex]{graphicx}
\usepackage{bm}
\usepackage[usenames,dvipsnames]{color}
\usepackage{mathtools}
\DeclarePairedDelimiter\bra{\langle}{\rvert}
\DeclarePairedDelimiter\ket{\lvert}{\rangle}
\DeclarePairedDelimiterX\braket[2]{\langle}{\rangle}{#1 \delimsize\vert #2}
\usepackage{amsmath}

\DeclareMathOperator \sh {sh}
\DeclareMathOperator \ch {ch}
\newcommand{\bg}{ \begin{gather} }
\newcommand{\eg}{\end{gather}}
\newcommand{\be}{ \begin{equation} }
\newcommand{\ee}{\end{equation}}
\newcommand{\bea}{ \begin{eqnarray} }
\newcommand{\eea}{\end{eqnarray}}

\newcommand{\str}{\mathop{\rm Str}}

\begin{document}

\title{Fractality of wave functions on a Cayley tree:\\ Difference between a tree and a locally tree-like graph without boundary}

\author{K.\,S.~Tikhonov}
\affiliation{L.\,D.~Landau Institute for Theoretical Physics RAS, 119334 Moscow, Russia}
\affiliation{Institut f{\"u}r Nanotechnologie, Karlsruhe Institute of Technology, 76021 Karlsruhe, Germany}
\affiliation{Condensed-matter Physics Laboratory, National Research University Higher School of Economics, 101000 Moscow, Russia}

\author{A.\,D.~Mirlin}
\affiliation{Institut f{\"u}r Nanotechnologie, Karlsruhe Institute of Technology, 76021 Karlsruhe, Germany}
\affiliation{Institut f{\"u}r Theorie der Kondenserten Materie, Karlsruhe Institute of Technology, 76128 Karlsruhe, Germany}
\affiliation{L.\,D.~Landau Institute for Theoretical Physics RAS, 119334 Moscow, Russia}
\affiliation{Petersburg Nuclear Physics Institute,188300 St.\,Petersburg, Russia.}

\begin{abstract}
We investigate analytically and numerically eigenfunction statistics in a disordered system on a finite Bethe lattice (Cayley tree). We show that the wave function amplitude at the root of a tree is distributed fractally in a large part of the delocalized phase. The fractal exponents are expressed in terms of the decay rate and the velocity in a problem of propagation of a front between unstable and stable phases. We demonstrate a crucial difference between a loopless Cayley tree and a locally tree-like structure without a boundary (random regular graph) where extended wavefunctions are ergodic.
\end{abstract}

\maketitle
\section{Introduction} 
\label{s1}

Anderson localization~\cite{anderson58}---one of most fundamental and ubiquitous quantum phenomena---remains in the focus of current experimental and theoretical research. Of particular interest are Anderson transitions between delocalized and localized phase \cite{evers08}. A disordered quantum system can be driven through such a transition by changing one of control parameters such as, e.g., disorder strength or energy.  For a conventional situation in $d$ spatial dimensions, an analytical study of the transition requires approximations (such as the $\epsilon$-expansion in $d=2+\epsilon$ dimensions). Remarkably, for models on the Bethe lattice (a tree with constant connectivity) the problem of the Anderson transition allows for an exact solution, making it possible to establish the transition point and the corresponding critical behavior \cite{abouchacra73,efetov85,zirnbauer86,efetov87,verbaarschot88,mirlin91}.

It is worth emphasizing that the analysis in Refs.~\onlinecite{abouchacra73,efetov85,zirnbauer86,efetov87,verbaarschot88,mirlin91} was performed on an infinite Bethe lattice. This formulation is appropriate for determination of the position of the transition point and for investigation of properties of localized wave functions and of finite-frequency correlation functions in the delocalized phase. The obtained results are also valid (up to small corrections) for a finite system, assuming the number of sites $N$ is sufficiently large. For correlation functions in the delocalized phase, the precise condition on $N$ depends on the frequency and on the distance to the transition point. This means that, for given parameters of the problem and for a given frequency $\omega$, there is a certain characteristic size $N_\omega$ such that for $N \gg N_\omega$ the correlation functions are essentially independent on $N$, i.e. the system can be considered as infinite. In this situation, the correlation functions are also independent on boundary conditions. The physical reason for this independence of finite-frequency correlation functions on $N$ and on boundary conditions is quite transparent. The frequency $\omega$ sets a characteristic spatial scale $L_\omega$ (which is a typical displacement of a particle in a time $\sim 1/\omega$). Once a ``linear size'' ($\sim \ln N$)  of the lattice becomes much larger than $L_\omega$, the system becomes effectively infinite and the boundary conditions do not play a role, since the particle simply has no time to find out what is the system size and the boundary conditions. 

 There is a class of important observables, however, for which the situation is more intricate. 
These include the statistics  of eigenfunctions and energy levels on the delocalized side of the transition. 
 Contrary to finite-frequency correlation functions, such observables simply cannot be defined on an infinite lattice, i.e., their mere definition requires a consideration of a finite system.   On the other hand, for a Bethe lattice of finite size (also known as Cayley tree) most sites are on the boundary (at variance with finite-$d$ problems). Thus, one can expect (and we will show in this paper that this expectation is correct) that the presence of boundary may affect the wave function and level statistics in the delocalized phase in a crucial way. 
 
In view of expected influence of the boundary, it is natural to consider  a modification of the model that allows one to eliminate boundary effects. One such generalization is provided by the  sparse random matrix (SRM) ensemble (known in mathematical literature as Erd\"os-R\'enyi graphs) studied analytically in Ref.~\onlinecite{sparse}. Another, closely related, possibility, is to consider a random regular graph (RRG), which is essentially a finite portion of Bethe lattice wrapped onto itself. The RRG and SRM ensembles are very similar tree-like models without boundary (and with loops of typical size $\sim \ln N$). The difference between them (in the connectivity being fixed in RRG and fluctuating around its average value in SRM) is immaterial for our discussion. These ensembles can be viewed as describing a tight-binding model on a lattice that has locally a tree-like structure but does not possess a boundary. It was found in Ref.~\onlinecite{sparse} that in the delocalized phase and in the limit of large number of sites $N$ (i) the level statistics takes the Wigner-Dyson form, and (ii)
 the inverse participation ratio (IPR) $P_2=\sum_i|\psi_i|^4$ characterizing fluctuations of an eigenfunction $\psi$ on the infinite cluster (with $\psi_i$ being the wavefunction amplitude on site $i$) scales with $N$ as $P_2\simeq{C/N}$. 
Here the prefactor $C(W)$ depends on the disorder strength $W$, approaching its Gaussian-ensemble value 3 deeply in the metallic phase ($W \to 0$) and diverging as $ \ln C \propto (W_c-W)^{-1/2}$ at the localization transition ($W=W_c$). 
Numerical results of Refs.~\onlinecite{Sade03,slanina12} for the model on random regular graphs supported the transition from the Poisson to the Wigner-Dyson statistics at the Anderson transition.

In recent years, the Anderson localization on RRG has attracted a renewed attention, in particular, in view of its connections with the problems of many-body localization in quantum dots \cite{sivan94a,sivan94,altshuler97,jacquod97,mirlin97,silvestrov97,silvestrov98,DLS01,mejia-monasterio98,leyronas99,weinmann97,berkovits98,leyronas00,rivas02,gornyi16,Kozii16}
and in spatially extended systems with localized single-particle states and with short-range
\cite{fleishman80,gornyi05,basko06,ros15,oganesyan07,monthus10,bardarson12,serbyn13,gopalakrishnan14,luitz15,nandkishore15,karrasch15,agarwal15,barlev15,gopalakrishnan15,reichmann15,lerose15,feigelman10,serbyn15,vosk15,ACP15,knap15,ovadyahu,ovadia15,schreiber15,bordia15,Bera15,Geraedts16,bloch16}   or long-range  \cite{Burin98,Demler14,Burin15,Smith16,Gutman16} interactions.
Biroli et al. \cite{biroli12} explored the level and eigenfunction statistics in the RRG model on the  delocalized side of the transition (disorder $W$ smaller than the critical disorder $W_c$). 
It is well understood that for conventional disordered systems (i.e., in a finite spatial dimensionality $d$) the level and eigenfunction statistics have three distinct types of behavior at the localized, critical, and delocalized fixed points \cite{evers08,mirlin00}. These statistics have been thus efficiently used to locate the Anderson transition and to study the associated critical behavior \cite{shklovskii93,Hofstetter93,Zharekeshev95,Varga95,Kaneko99,Milde00,Rodriguez09}.
The authors of Ref.~\onlinecite{biroli12}  interpreted the data for matrix sizes $N$ between 512 and 8192 as a possible indication of the intermediate ``non-ergodic delocalized'' phase (between the conventional delocalized phase and the localized phase, i.e. in a disorder range $W_T < W < W_c$ with a certain $W_T$). They argued that this phase is characterized (in the limit $N\to\infty$) by Poisson level statistics and by the IPR that does not scale as $1/N$. Subsequently, the problem of Anderson localization on RRG graphs was considered numerically by De Luca et al. \cite{deluca14}. These authors focussed on the eigenfunction statistics for systems with $N$ in the range from 2000 to 16000.  On this basis, they conjectured that eigenstates are multifractal in the whole delocalized phase, i.e., for all $0 < W < W_c$. This would imply, in particular, that the IPR scales in the large-$N$ limit as $P_2 \propto N^{-\mu}$ with the exponent $\mu(W)$ satisfying $\mu(W) < 1$ for all $W < W_c$.  

Clearly, the conclusions of Refs.~\onlinecite{biroli12,deluca14} based on numerical data are in conflict with the analytical predictions of Ref.~\onlinecite{sparse}.  This apparent contradiction was resolved in a recent work of the present authors with Skvortsov \cite{tms16}. In that work, we performed a numerical investigation of level and eigenfunctions statistics on the delocalized side of the Anderson transition on RRG, for system sizes $N$ from 512 to 262144. Our results fully support the analytical prediction of Ref.~\onlinecite{sparse} that states in the delocalized phase are ergodic in the sense that their IPR scales as $1/N$ and their level statistics is of Wigner-Dyson form in the limit $N\to \infty$.
We showed that the data can be interpreted in terms of a finite-size crossover from relatively small ($N\ll N_c$) to large ($N\gg N_c$) system, where  $N_c$ is the correlation volume diverging exponentially at the transition. More specifically, numerically found values of $N_c$ are in agreement with the analytical prediction\cite{sparse} $\ln N_c \sim (W_c-W)^{-1/2}$.
A distinct feature of this crossover is a pronounced non-monotonous behavior of observables as functions of $N$ on the delocalized side of the Anderson transition. This non-monotonicity has a profound origin in the nature of the Anderson-transition fixed point for a tree-like structure (or, equivalently, in the limit $d\to\infty$). Specifically, for $N\ll N_c$ the system flows towards the Anderson-transition fixed point which has on RRG properties analogous to the localized phase. Only when $N$ exceeds $N_c$, the flow changes direction and the system starts to approach its $N\to\infty$ ergodic behavior. The  non-monotonous behavior, in combination with exponentially large values of $N_c$, makes the finite-size analysis highly non-trivial: taking data in a limited range of $N$ may mislead one to a conclusion that the system is ``non-ergodic'' in the delocalized phase. 

Thus the analytical theory of the delocalized phase on RRG\cite{sparse} (ergodicity manifesting itself on scales $N\gg N_c$) is now supported by numerics\cite{tms16}. On the other hand, properties of delocalized eigenfunctions on a finite Cayley tree have remained largely unexplored. Several years ago, Monthus and Garel \cite{monthus11} studied numerically the statistics of transmisson amplitudes on a Cayley tree in the Miller-Derrida scattering geometry and concluded that it has a multifractal form in the delocalized phase. This suggest that eigenfunctions of an isolated Cayley tree may also have peculiar properties. In fact, some indications of this were obtained in an earlier paper by the same authors \cite{monthus09}.

In the present paper we show that wave functions in the Cayley-tree problem have indeed very unusual properties. We study, both analytically and numerically, the statistics of eigenfunctions in the root of a (finite) Cayley tree on the delocalized side of the Anderson transition. We show that, in stark contrast to ergodicity of delocalized states on RRG, the eigenfunctions on a tree show a fractal behavior in a large part of the delocalized phase.

\section{Wavefunction statistics at the root of Caylee tree: Analytical approach}
\label{s2}

\subsection{Model}

We study non-interacting spinless fermions hopping over a Cayley tree  with connectivity $K = m+1$  in a potential disorder,
\begin{equation}
\label{H}
\mathcal{H}=t\sum_{\left<i, j\right>}\left(c_i^+ c_j + c_j^+ c_i\right)+\sum_{i=1} \varepsilon_i c_i^+ c_i\,,
\end{equation}
where the sum is over the nearest-neighbour sites of the Cayley tree. The energies $\varepsilon_i$ are independent random variables sampled from a uniform distribution on $[-W/2,W/2]$. The investigation of this model was pioneered by Abou-Chacra et al, Ref.~\onlinecite{abouchacra73}; its solution in the framework of supersymmetry approach was obtained in Ref.~\onlinecite{mirlin91}.  It is useful to consider also an $n$-orbital generalization of the problem (with $n\gg 1$) which can be viewed as describing an electron hoping between metallic granules located at the nodes of the same Cayley tree. The Hamiltonian of such a granular system reads
\begin{equation}
\label{HG}
\mathcal{H}=t\sum_{\left<i, j\right>}\sum_{p,q=1}^n \left(c_{ip}^+ c_{jq} + c_{jq}^+ c_{ip}\right)+\sum_{i}\sum_{p=1}^n \varepsilon_{ip} c_{ip}^+ c_{ip}\,.
\end{equation}
For large $n$, the $n$-orbital problem can be mapped onto a supersymmetric $\sigma$-model \cite{efetov85,zirnbauer86,efetov87,verbaarschot88}. While the $n=1$ Anderson model and its $n\gg 1$ generalization ($\sigma$-model) turn out to exhibit the same gross features, analytical calculation are somewhat simpler within the $\sigma$-model. For this reason, we find it instructive to carry out the analysis first within the $n\gg 1$ model (i.e., the $\sigma$-model). Later we will return to the $n=1$ Anderson model and discuss corresponding modifications. 

\begin{figure}
\centering
\includegraphics[width=.25\textwidth]{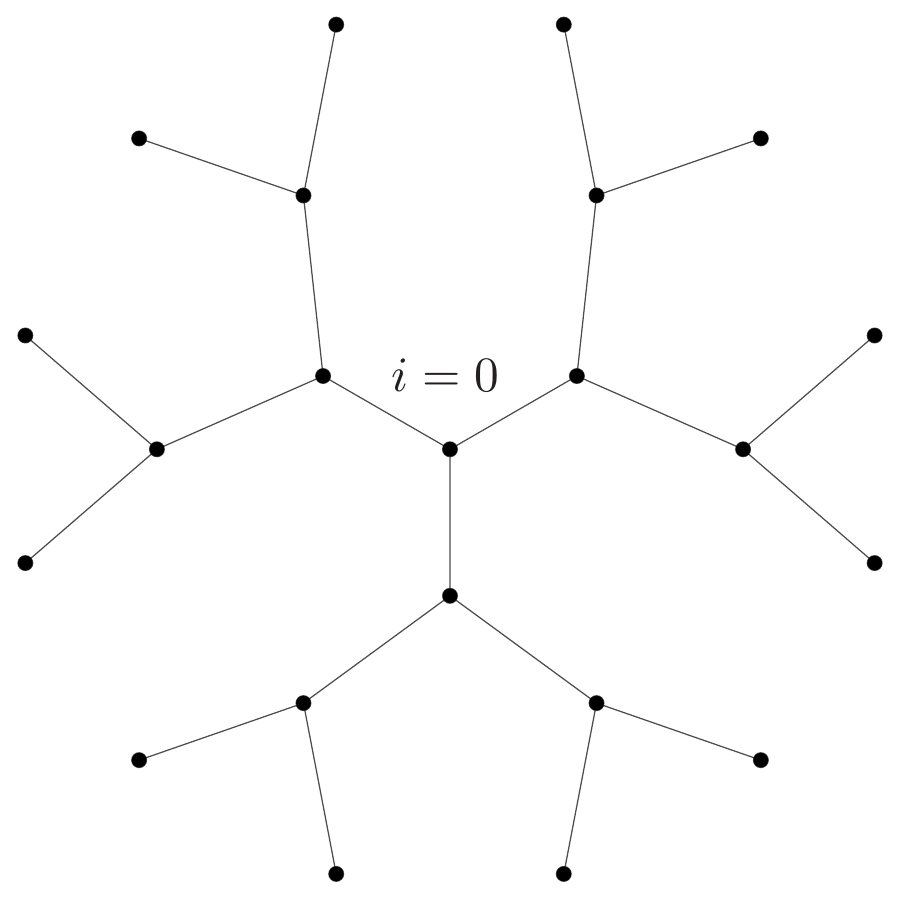}
\caption{Cayley tree with branching number $m=2$ and $s_0=3$ generations. In this paper we study the eigenfunction statistics at the root of the tree, $i=0$.}
\label{fig_ct}
\end{figure}

\begin{figure}
\centering
\includegraphics[width=.25\textwidth]{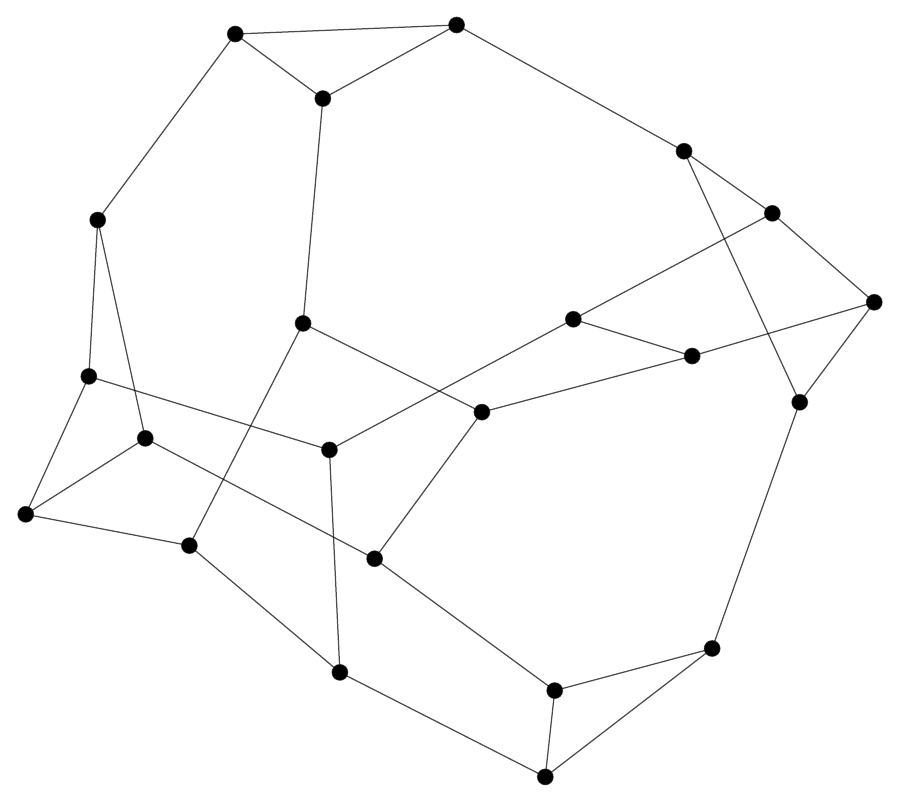}
\caption{Random regular graph with the same connectivity ($m=2$) and the same number of vertices ($N=22$) as the Cayley tree in Fig.~\ref{fig_ct}.}
\label{fig_rrg}
\end{figure}

We will assume free boundary conditions (which means that we consider an isolated system) and study statistics of wavefunction amplitudes $u_i=|\psi_i |^2$. Contrary to RRG, the sites of a Cayley tree are clearly not equivalent, and distribution function of $u_i$ depends on the distance from the site $i$ to the boundary of the tree. For simplicity, we fix $i$ to be at the root, $u\equiv u_{0}$ and study the distribution function $\mathcal{P}_N(u)$ as a function of both $u$ and the size of a tree $N$. As an example, in Figs.~\ref{fig_ct} and \ref{fig_rrg} we show a Cayley tree with the connectivity $m=2$  and $s_0=3$ generations and a representative of the RRG ensemble withe the same connectivity and the same number of vertices, $N=22$.

A convenient tool to explore analytically the eigenfunction statistics in a non-interacting disordered system is the supersymmetry method. The moments of wavefunction amplitudes  can be expressed in terms of Green functions $G_{R(A)}$ at coinciding points (in our case, at the root) as follows \cite{mirlin00}:
\be
\label{def}
\left<|u|^q\right> = \frac{i^{q-2}}{2\pi\nu N} \lim_{\eta\to 0} (2\eta)^{q-1}\left<G_R^{q-1}G_A\right>
\ee
with
\be
G_{R(A)}=\bra{0}\left(\varepsilon-\hat H\pm i\eta\right)^{-1}\ket{0}.
\label{Green}
\ee
Here $\hat H$ is the single-particle Hamiltonian, $\varepsilon$ the energy, and $\nu$ the density of states (at energy $\varepsilon$). 

\subsection{Sigma model}

Let us start with the case of $n\gg 1$ orbitals per each lattice site as described by Eq.~(\ref{HG}). In this situation, the theory can be reduced \cite{efetov85} to the supersymmetric $\sigma$-model with the action
\be
\label{action}
S[Q]= - J\sum_{\left<i, j\right>}\str(Q_i-Q_j)^2 + \frac{\pi\eta}{2\delta_0}\sum_i \str(\Lambda Q_i).
\ee
Here $Q_i$ are $8 \times 8$ supermatrices satisfying the condition $Q^2=1$, the symbol $\str$ denotes the supertrace (defined as trace of the boson-boson block minus trace of the fermi-fermi block), $\delta_0 = \nu^{-1} = W/n$ is the mean level-spacing on a granule, and $J=\left(t/\delta_0\right)^2$ is the dimensionless coupling constant. The microscopic model (\ref{HG}) belongs to the orthogonal (AI) symmetry class, determining the corresponding symmetry of the  $\sigma$ model. When the time-reversal symmetry is broken (e.g., all hopping amplitudes $t_{ij}^{pq}$ are complex with random phases), the symmetry class becomes unitary (A). The physics that we discuss in this paper is essentially the same in both cases. Since the unitary-symmetry case is somewhat simpler technically, we will focus on it below for the sake of transparency of exposition. In this case, $Q_i$ in Eq.~(\ref{action}) become $4\times 4$ supermatrices, and the action  (\ref{action}) acquires an additional overall factor of two.

The average product of Green functions in Eq.~(\ref{def}) can be represented as a sigma-model correlation function of the following form\cite{mirlin00}:
\bea
\label{srep}
\left<|u|^q\right> &=& -\frac{q}{2N}\lim_{\eta\to 0}(2\pi\eta/\delta_0)^{q-1} \nonumber \\
& \times& \int DQ \:Q^{q-1}_{0;11,bb}Q_{0;22,bb}\:e^{-S[Q]},
\eea
where the preexponential factor depends only on the matrix $Q_0$ at the root of the Cayley tree (which is the point where we study the eigenfunction statistics). The first two indices of $Q$ correspond to the advanced-retarded and the last two to the boson-fermion decomposition. 

To evaluate the functional integral in Eq.~(\ref{srep}), it is convenient first to integrate out all degrees of freedom $Q_i$ with $i\ne 0$ and, at the last step, to take the integral over the matrix $Q_0$ associated with the root of the tree. The tree structure of the lattice greatly simplifies the task: the matrices $Q_i$ can be consecutively integrated out, starting from the boundary (the ``leaves'' of the tree) and proceeding layer by layer towards the root. This iterative procedure can be described in terms of functions $\Psi_s(Q)$ (with $s=0,1,2,\ldots$) defined in the following way. Consider a site $i$ of the Cayley tree. Consider one of $m$ branches of the tree that start at this site and do not contain the root. Perform the integration over the variables $Q_j$ associated with this branch, with the corresponding part of the weight  $e^{-S[Q]}$. Clearly, the result depends only on the matrix $Q_i$ (since the action does not couple sites on the branch to any sites of the remaining lattice other than $i$), and we denote it $\Psi^{(i)}(Q_i)$. In view of the symmetry of the Cayley tree, the function $\Psi^{(i)}$ will be identically the same for all sites $i$ located on a given distance $s$ from the boundary. (Here $s=0$ corresponds to leaves of the tree, $s=1$ to sites separated by one link from the boundary, etc.) Hence, the $\Psi$ functions can be naturally labeled by an index $s$, yielding a sequence of functions  $\Psi_s(Q)$ with $s=0,1,2,\ldots$.

It is easy to see that the functions $\Psi_s(Q)$ satisfy the following recurrence relation:
\be
\label{recurse}
\Psi_{s+1}(Q)=\int e^{-\str\left[-2J(Q-Q')^2+\frac{\pi\eta}{\delta_0}\Lambda Q'\right]}\Psi_{s}^m(Q')DQ',
\ee
with the initial condition $\Psi_0(Q) = 1$ at the boundary. After 
\be
s_0 = \frac{\ln N}{\ln m}
\label{s0}
\ee 
iterations of this recurrence relation, we obtain the function 
$\Psi_{s_0}(Q) $ at the root. 
To proceed further, we note that, in view of the symmetry of the $\sigma$-model action, the functions $\Psi_s(Q)$ in the unitary symmetry case depend only on two variables $1\leq\lambda_1<\infty$ and $-1\leq\lambda_2\leq 1$, which are the eigenvalues of the retarded-retarded block of the matrix $Q$, see Ref.~\onlinecite{mirlin00}. The variables $\lambda_1$ and $\lambda_2$ correspond to the non-compact (hyperbolic) and compact (spherical) sectors of the $\sigma$-model coset space.  As we are interested in the limit of $\eta \to 0$ at fixed $N$ (and hence at fixed $s_0$), we can further simplify the equation (\ref{recurse}). Specifically, in this limit only the dependence on $\lambda_1$ persists:
\be
\Psi_s(Q)\equiv\Psi_s(\lambda_1,\lambda_2) \to \Psi^{(a)}_s(2\pi\eta\lambda_1/\delta_0),
\label{asymptotic}
\ee
where the superscript $(a)$ indicates that we are dealing with the asymptotic, small-$\eta$ form of the function $\Psi_s$.

As is clear from Eq.~(\ref{srep}), the distribution function of the wave function intensity $u_0$ at the root is fully determined by the 
asymptotic form of the function 
\be
\label{Y}
Y(Q_0) = \Psi_{s_0}^{m+1}(Q_0)
\ee
 resulting from integrating out all degrees of freedom on the tree except for the matrix $Q_0$ at the root. Specifically, evaluating the integral over $Q_0$, one gets\cite{mirlin00}
\be
\mathcal{P}(u)=N^{-1}\partial_u^2 Y^{(a)}(u),
\label{Pu}
\ee
where $Y^{(a)}(u) = [\Psi^{(a)}_{s_0}(u)]^{m+1}$.

Let us emphasize that the order of limits (first $\eta\to 0$ at fixed $N$, after which arbitrarily large $N$ can be considered) is of crucial importance for properly extracting the eigenfunction statistics. We will return below several times to this important point and related issues. 

In the $\eta\to 0$ limit, in which the functions $\Psi_s$ depend on a single scalar variable [see Eq.~(\ref{asymptotic})], the recurrence relation (\ref{recurse}) can be substantially simplified. It is convenient to introduce $t = \ln(2\pi\eta\lambda_1/\delta_0)$ and to perform the change of variable 
\be
\label{Phi}
\Psi_s^{(a)}(e^t) = \Phi_s(t).
\ee 
One gets then the asymptotic recurrence relation
\be
\label{recsimple}
\Phi_{s+1}(t)=\int L(t-t')e^{-e^{t'}}\Phi^m_s(t')dt',
\ee
where the kernel $L(t)$ is given by 
\be
L(t)=\frac{2g\ch g+(2g\ch t-1)\sh g}{2\sqrt{2\pi g}}e^{-t/2-g\ch t},
\ee
with $g=J/8$.

\begin{figure}
\centering
\includegraphics[width=.5\textwidth]{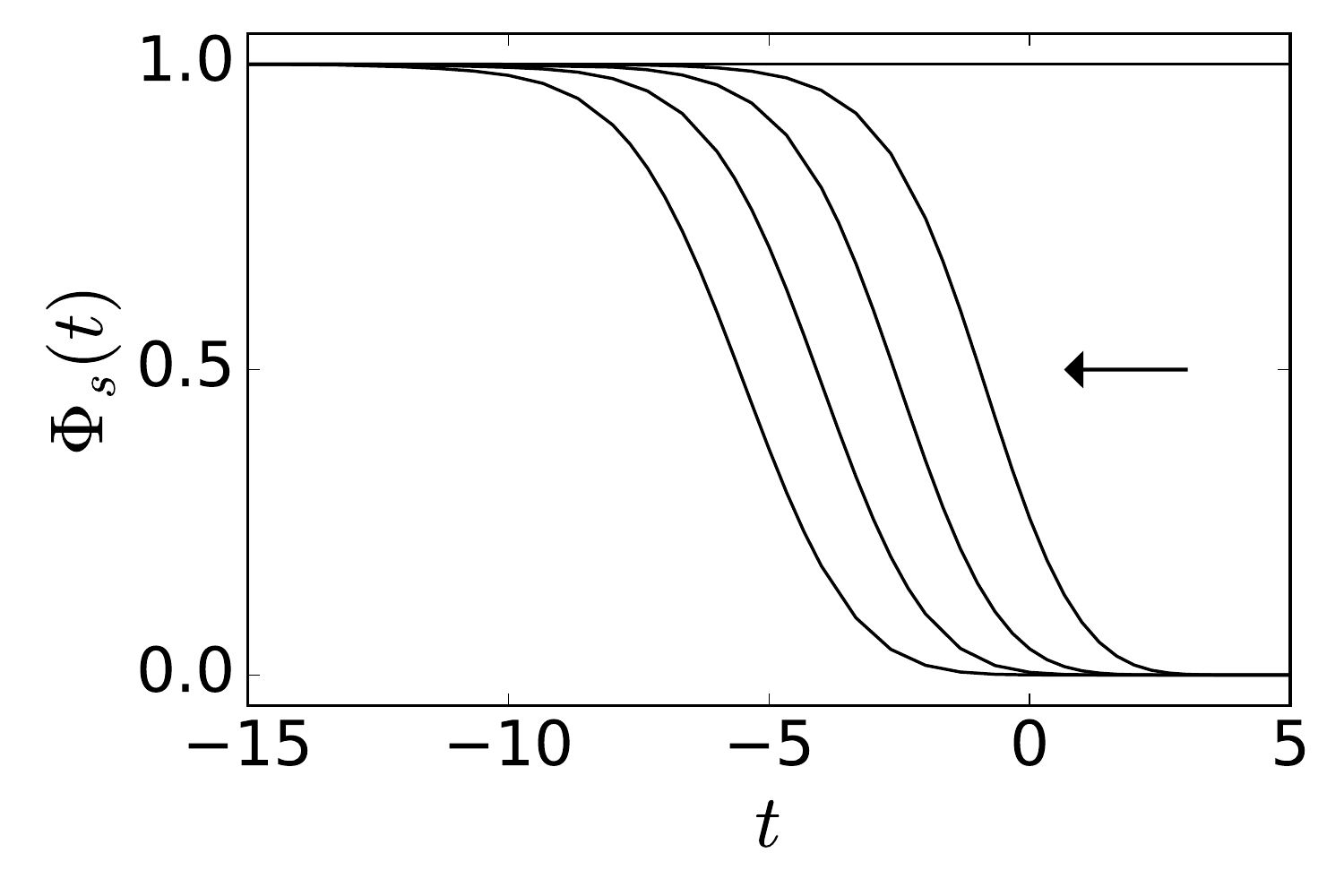}
\caption{Evolution of the kink $\Phi_s(t)$ on iterating the asymptotic recurrence relation Eq. (\ref{recsimple}) deeply in the delocalized phase ($g=1$) for $s=0,1,2,3,4$. }
\label{evolution}
\end{figure} 

Equation (\ref{recsimple}) was obtained by Efetov \cite{efetov85} and Zirnbauer\cite{zirnbauer86} in course of the analysis of the stability of the insulating phase with respect to the symmetry breaking perturbation (the term in the action proportional to $\eta$). It is useful to remind the reader about the essence of this analysis. For small $g$ (in the localized phase), the recurrence relation (\ref{recsimple}) yields a kink that stabilises after a few iterations. This means that the asymptotic self-consistency equation (obtained from Eq.~(\ref{recsimple}) by setting $\Phi_{s+1}=\Phi_s\equiv\Phi$) has a non-trivial solution $\Phi(t)$. Such a solution corresponds to the function $\Psi(Q)$ depending on $\lambda_1$ only on the scale $\lambda_1 \sim \eta^{-1}\delta_0$ set by the symmetry breaking term. The function $\Psi(Q)$ can be viewed as an order-parameter function, and the fact  that deviates from unity only on the scale set by $1/\eta$ corresponds to the absence of symmetry breaking. This is the characteristic feature of the localized phase.  On the other hand, for sufficiently large $g$ (delocalized phase), the drift of the kink, as described by the asymptotic recurrence relation  (\ref{recsimple}), continues indefinitely, 
see Fig.~\ref{evolution}. This
signifies the absence of a non-trivial solution of the asymptotic self-consistency equation and thus an instability of the localized phase.
The self-consistency equation corresponding to the general recurrence relation (\ref{recurse}) does have a solution which has a form of a kink with a position independent of $\eta$ for small $\eta$. Thus, the delocalized phase is characterized by broken symmetry from this point of view. Let us stress, however, that this self-consistent solution is not related to the problem we are considering in this work, since it corresponds to the opposite order of limits $N\to\infty$ and $\eta\to 0$. Specifically, to reach the stable solution by the iterative procedure (\ref{recurse}), one should consider the limit $N\to \infty$ at fixed (although small) $\eta$.  On the other hand, our problem of eigenfunction statistics on a finite Cayley tree is described by an opposite procedure: we should consider the limit $\eta\to 0$ at fixed (although large) $N$. In this situation, we are always in the range of applicability of the asymptotic recurrence relation (\ref{recsimple}) which describes, in the delocalized phase, a drift of the kink without saturation. 

In order to study the evolution of the kink, we consider Eq.~(\ref{recsimple}) in the region of $t<0$ and sufficiently large $|t|$ (on the left side of the front in Fig.~\ref{evolution}) where deviations from the ``localized'' value $\Phi(t)=1$ are small.  In this region, one can linearise Eq.~(\ref{recsimple}) in 
$\delta\Phi_s(t) = 1-\Phi_s(t)$ and drop the factor $e^{-e^{t'}}$. 
This yields 
\be
\label{recsimple2}
\delta\Phi_{s+1}(t)=m\int L(t-t')\,\delta\Phi_s(t')\,dt'.
\ee
Thanks to translational invariance of the kernel, the eigenfunctions of the integral operator $\hat L$ in the right-hand side of this equation 
are of the form $\psi_\beta(t)=e^{\beta t}$. The corresponding eigenvalues can be readily found:
\be
\label{specUnitary}
\epsilon_{\beta}=\frac{2g K_{\beta+1/2}(g)\sh g + 2K_{\beta-1/2}(g)(g\ch g-\beta\sh g)}{\sqrt{2\pi g}}.
\ee

\begin{figure}
\centering
\includegraphics[width=.5\textwidth]{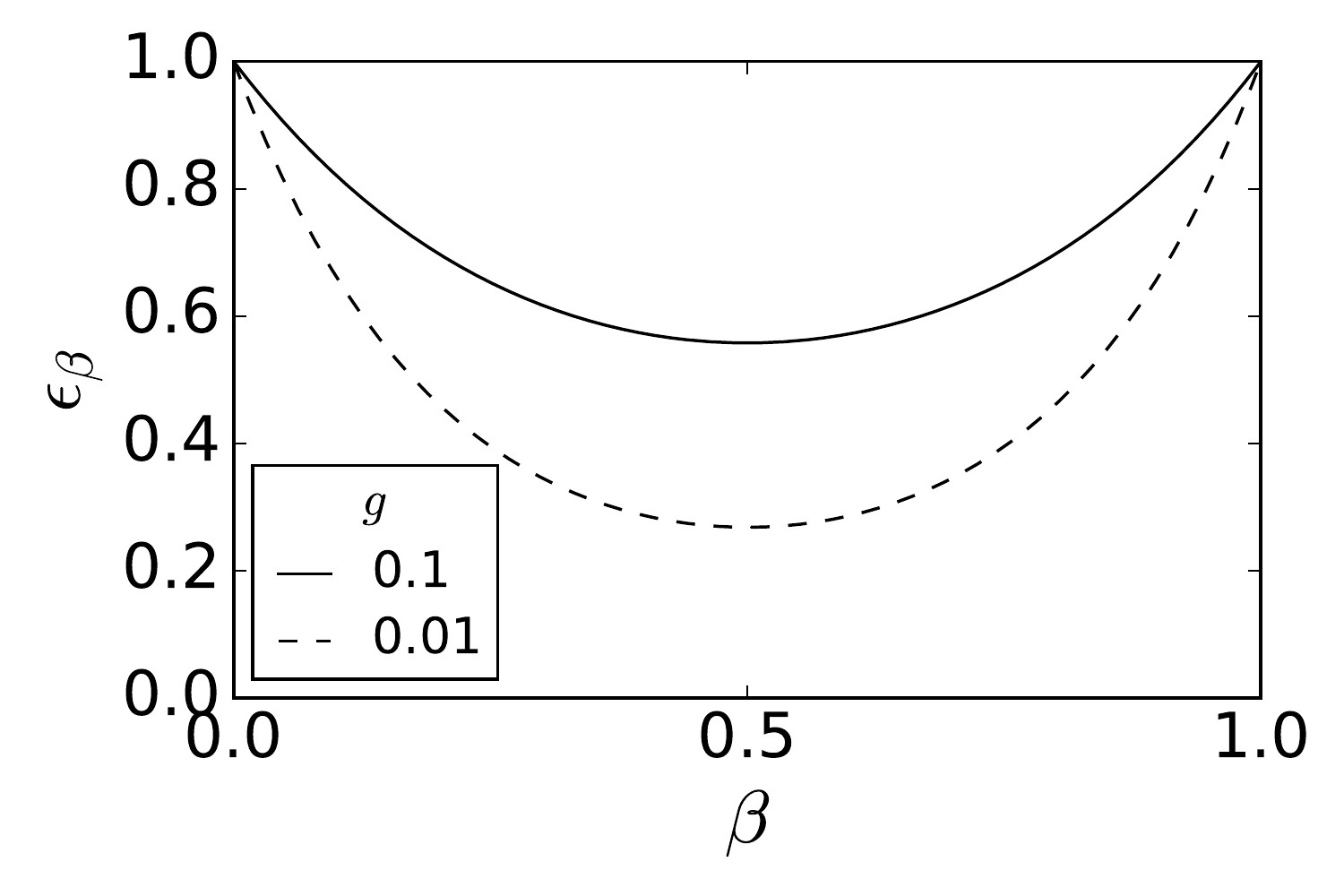}
\caption{Eigenvalue $\epsilon_\beta$ of the kernel $\hat L$ entering the recurrence relation (\ref{recsimple}) and its linearized form (\ref{recsimple2}), as a function of the exponent $\beta$, for two values of the coupling $g$. }
\label{eigen}
\end{figure} 

The function $\epsilon_\beta$ is shown in Fig.~\ref{eigen} for two values of the coupling  $g$.  Since $\delta\Phi\to 0$ at $t\to-\infty$, only $\beta >0$ are allowed. As this decay can not be faster than $e^t$ in view of the form of the symmetry-breaking term in Eq.~(\ref{recsimple}), $\beta$ also satisfies $\beta \le 1$. As we discuss below, the relevant values of $\beta$ satisfy $1/2 \le \beta \le 1$. This follows from the fact that the function $\epsilon_\beta$ (with $0 < \beta < 1$) has the following properties: (i) $\epsilon_1 = 1$, (ii) $\epsilon_\beta$ increases monotonously on the interval $1/2\le \beta\le 1$, (iii) $\epsilon_\beta= \epsilon_{1-\beta}$. These features (which imply that $\epsilon_\beta$ takes its minimum value at $\beta=1/2$) are clearly seen in Fig.~\ref{eigen}.

Before turning to the investigation of the delocalized phase, it is instructive to briefly recall the implications of the above form of the function $\epsilon_\beta$ for the analysis of stability of the localized phase\cite{efetov85,zirnbauer86}.
The system is in the localized phase if there exists a stationary solution of Eq.~(\ref{recsimple}). If this solution is characterized, in the asymptotic range of negative $t$, by an exponent $\beta$, then the following condition should be fulfilled according to Eq.~(\ref{recsimple2}):
\be
m \epsilon_{\beta} = 1.
\label{lambda-loc}
\ee
In the limit of infinitely strong disorder $g\to 0$, this is fulfilled for $\beta=1$. For $g>0$ one has $m \epsilon_1 > 1$, so that the initial perturbation $\delta\Phi(t) \propto e^t$ imposed by the symmetry-breaking term in Eq.~(\ref{recsimple}) grows according to the linear equation Eq.~(\ref{recsimple2}), and the kink starts moving to the left. For not too small $t$, this growth is, however, slowed down by non-linear effects  in Eq.~(\ref{recsimple}). As a result, the exponent $\beta$ characterizing the dependence $\delta\Phi(t) \propto e^{\beta t}$ becomes smaller, and the kink moves slower. For sufficienty small $g$, this process stops when such $\beta$ from the interval $1/2<\beta<1$ is reached that Eq.~(\ref{lambda-loc}) is satisfied. In this situation, a stationary solution of Eq.~(\ref{recsimple}) is reached, i.e., the system is in the localized phase. It follows that the critical value $g_c$ of the coupling $g$ corresponding to the Anderson transition between localized and delocalized phases is determined by the equation \cite{efetov85,zirnbauer86}
\be
m\epsilon_{1/2}=1.
\label{critical-g}
\ee

We are now ready to begin the analysis of the delocalized regime, $g>g_c$, when no stationary solution exists. In view of the above arguments, we expect that, after some  evolution on a short time scale, the solution $\Phi_s$ will reach a steady regime characterized by a certain exponent $\beta$ and a drift velocity $v_\beta$. The reduction of $\beta$ in comparison to its initial value $\beta = 1$ has the same origin---slowing down by non-linear terms---as in the localized phase. 
Assuming $\delta\Phi_i(t)\propto e^{\beta t}$, we get from Eq.~(\ref{recsimple2}) the following form of the solution on the leading edge of the front:
\be
\delta\Phi_s(t)= c e^{\beta t+ s \ln m\epsilon_{\beta}}.
\label{delta-Phi}
\ee
Thus, the front moves to the left with the velocity
\be
\label{velocity}
v_\beta = \frac{\ln m\epsilon_{\beta}}{\beta}.
\ee
When speaking about the velocity, we consider the variable $t$ as playing a role of spatial coordinate, and the recurrence-relation step $s$ as a representing a time. Clearly, here these notions of ``space'' and ``time'' are fictitious and introduced only for terminological convenience. However, as we discuss below, the present problem has much in common with a variety of non-linear problems where the time and the space corresponding to our $s$ and $t$ are real. 

A question of central importance is how the exponent $\beta$, and thus the velocity $v_\beta$, Eq.~(\ref{velocity}), is selected. 
As discussed above, the non-linear terms in Eq.~(\ref{recsimple2}) reduce $\beta$ compared to its initial value $\beta=1$, thus reducing 
the velocity $v_\beta$. This may happen until the minimal possible velocity is reached. Thus, the value $\beta_*$ of the exponent $\beta$ characterizing the steadily moving kink is determined by the condition of minimal velocity:
\be
\label{lambda-opt1}
v_{\beta_*} < v_\beta  \qquad {\rm for} \ \ 1/2\le\beta\le1, \ \ \beta \ne \beta_*
\ee
As we show below, for a part of the delocalized phase this minimum is reached at the boundary, $\beta_*=1$, while for the rest of the delocalized phase $\beta_*$ is located strictly inside the interval,  $1/2 < \beta_* < 1$, and is thus determined by the equation
\be
\label{lambda-opt2}
\left. \frac{dv_\beta}{d\beta} \right|_{\beta= \beta_*} = 0. 
\ee
The arguments in favor of the selection of minimal velocity---out of those provided by the linear equation (\ref{recsimple2})---due to non-linearities in Eq.~(\ref{recsimple2}) can be supported by more formal analysis of stability of the solution. In fact, such an analysis is available in the literature in context of a broad class of problems that are closely related mathematically although have very different origin. Specifically, these problems deal with nonlinear equations describing propagation of a front between an unstable (in our case, $\Phi=1$) and stable (in our case, $\Phi=0$) phases. The simplest partial differential equation of this type is known as Fisher-KPP equation, as it was first introduced by Fisher \cite{fisher37} and by Kolmogorov, Petrovskii, and Piskunov\cite{kpp} in the context of propagation of advantageous genes. Later, it has been realized that similar problems of traveling waves in reaction-diffusion systems arise in a great variety of further areas, including, in particular, fluid dynamics, propagation of domain walls in liquid crystals, chemical reactions, bacterial growth, propagation of combustion fronts, etc., see Refs.~\onlinecite{derrida88,saarloos88,saarloos03,brunet15} and references therein. A connection between the problem of statistical properties of various observables in Anderson localization at Cayley tree and that of traveling wave propagation was emphasized in Ref.~\onlinecite{monthus09}. 
The stability analysis \cite{saarloos88} shows that the selected velocity is determined by so-called marginal stability condition, which exactly corresponds to minimization of drift velocity, i.e., minimization of $v_\beta$ in our notations.  

Thus,  $\beta_*$ and $v_*$  are determined by the minimum-velocity condition. Before turning to the evaluation of the dependence $\beta_*$ and $v_*$ on the coupling constant $g$, let us analyse how these quantities manifest themselves in the wave function statistics. We write the obtained function $\Phi_s(t)$ in the form
\be
\label{psires}
\Phi_s(t) \simeq \begin{cases}
\begin{array}{ll}
1 - e^{t+s\ln m}, \ \ \ & t \lesssim t_-; \\
1-c\, e^{\beta_* (t+s\alpha_* \ln m )}, \ \ \ &     t_- \lesssim t \lesssim  t_+   ;\\
0,\ \ \ &  t \gtrsim t_+\ ,
\end{array}
\end{cases}
\ee
where $c$ is a numerical constant and $t_{\pm}$ are defined in Eqs (\ref{t-min}), (\ref{t-max}). We have also introduced 
\be
\label{alphadef}
\alpha_*= \frac{v_{\beta_*}}{\ln m} = \frac{\ln\left(m\epsilon_{\beta_*}\right)}{\beta_*\ln m}.
\ee
The reason for introducing $\alpha_*$ (which, as we show below, satisfies $0<\alpha_*\le1$) according to Eq.~(\ref{alphadef}) will become clear momentarily.

The first line of Eq.~(\ref{psires}) corresponds to the very-far asymptotics of the kink, which is the range of $t$ where the nonlinear effects have not developed yet (within the given time $s$).  In this region the evolution of $\delta\Phi_s(t)$ is controlled  by the linearized equation (\ref{recsimple2}), in combination with the symmetry breaking term $e^{-e^{t'}}$. Since the corresponding eigenvalue of the operator $\hat L$ is $\epsilon_1 = 1$, the behavior of $\Phi_s(t)$ in this region is completely universal (e.g., is the same as it would be in the limit of infinite coupling $g\to\infty$ when the kernel $L(t)$ becomes a delta function and the whole system becomes a GUE ensemble) and is given by the first line of Eq.~(\ref{psires}). The boundary $t_-$ between this region of GUE-like behavior and the non-trivial regime of the type (\ref{delta-Phi}) characterized by $\beta_*<1$ [second line of Eq.~(\ref{psires})] is found from the corresponding matching condition,
\be
\label{t-min}
t_-+s\ln m = \beta_* (t_-+s\alpha_* \ln m ).
\ee

For $t$ to the right of the front, $t > t_+$ , where $\Phi_s(t) \ll 1$, Eq.~(\ref{recsimple}) implies a very fast (double exponential) decrease of  $\Phi_s$ with $t$. This region will not thus play any role in the following analysis of eigenfunction statistics, and we can safely replace $\Phi_s$ there by zero [third line of Eq.~(\ref{psires})]. The boundary $t_+$ is straightforwardly found by the matching condition,
\be
\label{t-max}
t_+ +s\alpha_* \ln m = 0.
\ee

Transforming the function $\Phi(s)$, Eq.~(\ref{psires}), into $\Psi^{(a)}(u)$ according to Eq.~(\ref{Phi}), substituting the result into Eq.(\ref{Pu}) and using Eq.~(\ref{s0}) for the number of iteration needed to reach the root, we find the distribution $\mathcal{P}(u)$ of eigenfunction intensity at the root. The non-trivial part of this distribution, which is of interest for us, corresponds to the  $t_- \lesssim t \lesssim  t_+ $ regime of Eq.~(\ref{psires}):
\be
\label{presult}
\mathcal{P}(u) \simeq 
c' N^{-1+\alpha_*\beta_*}u^{\beta_*-2},\ \ \ 
N^{-\frac{1-\alpha_* \beta_*}{1-\beta_*}}\lesssim u \lesssim N^{-\alpha_*},
\ee
where $c'$ is a numerical constant.
Thus, the eigenfunction distribution $\mathcal{P}(u)$ has a power-law form  in a parametrically broad range of $u$ if $\alpha_* < 1$.  The behavior of ${\cal P}(u)$ outside of this range of $u$ does not affect the moments: the distribution quickly vanishes at large $u > N^{-\alpha_*}$ and saturates at small $u < N^{-(1-\alpha_* \beta_*)/(1-\beta_*)}$.  It is easy to check that the normalization condition  $\int du \mathcal{P}(u)=1$ is satisfied; the dominant contribution to the normalization comes from the lower limit of the power-law behavior  (\ref{presult}). 

Having in our disposition the distribution function, we can easily find the scaling behavior of all moments 
\be 
P_q=N\left<\psi^{2q}\right> = N \int du \, u^q \mathcal{P}(u)
\label{IPR-def}
\ee
with the system size $N$. 
 We have included the factor $N$ in the definition (\ref{IPR-def}) of $P_q$ to make it analogous to the familiar definition of inverse participation ratios. 
 In the conventional (ergodic) delocalized phase, one has the scaling $P_q \propto N^{1-q}$.
 
 Let us start with the second moment $P_2=N\left<\psi_r^{4}\right>=N\int \mathcal{P}(u) u^2 du$. Using (\ref{presult}), we immediately see that the value of this integral is determined by the upper cutoff of the power-law behavior, $u \sim N^{-\alpha_*}$, yielding
 \be
 P_2 \propto N^{-\alpha_*}.
 \label{P2}
 \ee
 Thus, while for $\alpha_*=1$ we have a conventional $1/N$ scaling characteristic for a delocalized phase, in the case of $0<\alpha_*<1$ the scaling is of fractal character. We will show below that such a fractal scaling is realized in a large part of the delocalized phase in the considered Cayley tree model.  
 Let us emphasize that Eq.~(\ref{P2}) represents a true large-$N$ asymptotic behavior of the wave function moment. Thus, the model of a finite Cayley tree with a boundary is crucially different from the RRG model which shows ergodic behavior (in particular, $P_2 \propto 1/N$) in the limit $N\to \infty$ in the whole delocalized phase. We will return to this very important difference below.

\begin{figure}
\centering
\includegraphics[width=.5\textwidth]{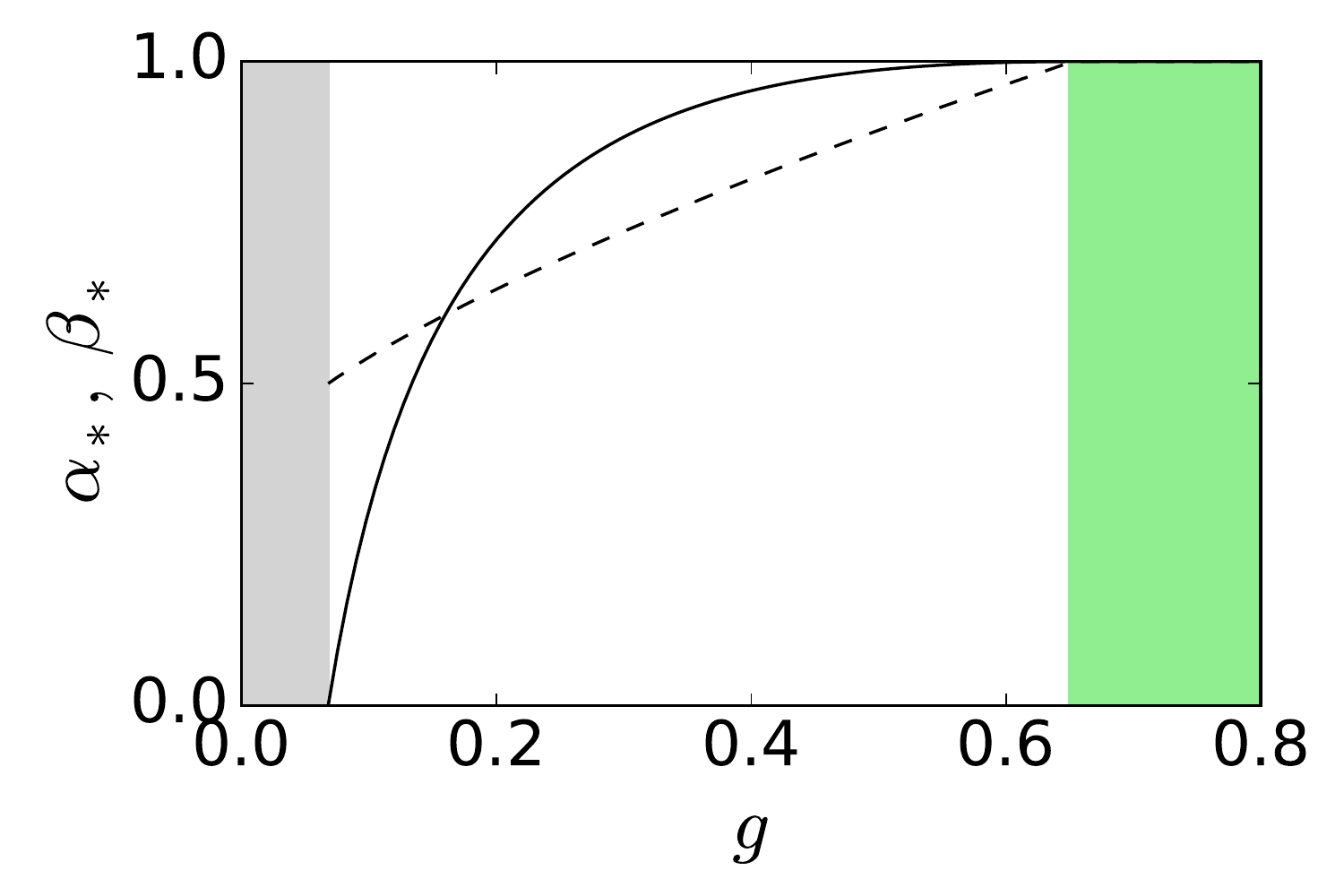}
\caption{Exponent $\beta_*(g)$ characterizing the drifting kink (dashed) and the fractal dimension $\alpha_*(g)$ (solid) for the sigma model of unitary symmetry on Cayley tree with $m=2$. Three phases are seen: (i) $g < g_c$ -- localized phase (gray), (ii) $g_c < g < g_e$ -- delocalized fractal (non-ergodic) phase with $1/2 < \beta_*<1$ and $0<\alpha_*<1$, and  (iii) $g > g_e$ -- delocalized ergodic phase (green) with $\beta_* = \alpha_* = 1$.  
 }
\label{figlambda}
\end{figure} 

Let us now evaluate $\beta_*$ and $\alpha_*$ as defined by Eqs.~(\ref{specUnitary}), (\ref{velocity}),  (\ref{lambda-opt1}), and (\ref{alphadef}). It turns out that the delocalized phase, $g > g_c$, is subdivided into two parts, see Fig.~\ref{figlambda} where the dependences $\beta_*(g)$ and $\alpha_*(g)$ are shown for $m=2$.  Specifically, for large values of the coupling, $g > g_e$, the minimal velocity (\ref{lambda-opt1}) is achieved at the boundary point, 
$\beta_*=1$. As follows from Eqs.~(\ref{velocity}) and (\ref{alphadef}), in this situation $\alpha_*=1$. Thus, the region $g>g_e$ is a conventional (ergodic) delocalized phase. On the other hand, for intermediate range of couplings, $g_c< g < g_e$, we find that $\beta_*$ is located strictly inside the interval  $1/2 < \beta < 1$ and is thus determined by Eq.~(\ref{lambda-opt2}). In this situation the minimal velocity is smaller than the one corresponding to $\beta=1$, so that  $0<\alpha_*<1$. Therefore, the intermediate region $g_c< g < g_e$ is a non-ergodic (fractal) delocalized phase.
The values of $g_c$ and $g_e$ depend on the Cayley tree connectivity $m$; for $m=2$ they are $g_c= 0.068$ and $g_e = 0.65$.  The picture remains, however, qualitatively the same for other values of $m$ as well, since it is determined by general properties of $\epsilon_{\beta}$ described below Eq.~(\ref{specUnitary}).

Having determined the dependences $\beta_*(g)$ and $\alpha_*(g)$, we return now to the analysis of the wave function statistics. Similarly to the above calculation of the second moment $P_2$, see  Eq.~(\ref{P2}), we can find, by using the distribution function (\ref{presult}), other moments $P_q$. 
It is easy to see that all moments with $q>q_*$, where
\be
q_*=1-\beta_*,
\ee
 are determined by the upper limit of the power-law behavior in Eq.~(\ref{presult}), 
$u \sim N^{-\alpha_*}$, while all moments with $q<q_*$ are determined by the lower limit, $u\sim N^{-(1-\alpha_* \beta_*)/(1-\beta_*)}$.
 As a result, we obtain the scaling
 \be
 P_q\propto N^{1-q-\Delta_q}
 \label{Pq}
 \ee
 with the anomalous exponents $\Delta_q$ given by
\be
\label{qres}
\Delta_q=\begin{cases}
\begin{array}{ll}
\displaystyle q\frac{(1-\alpha_*)\beta_*}{1-\beta_*}, & \qquad q<q_*\:; \\[0.3cm]
(1-q)(1-\alpha_*), & \qquad q>q_* \:.
\end{array}
\end{cases}
\ee
In the region $g>g_e$ we have $\alpha_*=1$, so that all the exponents  (\ref{qres}) vanish, $\Delta_q=0$, and Eq.~(\ref{Pq}) reduces to conventional (ergodic) scaling of wave function moments in a delocalized phase. On the other hand, in the intermediate phase, $g_c < g < g_e$, where $\alpha_* < 1$, all $\Delta_q$ (with $q \ne 0,1$) are non-zero, i.e., all moments show a fractal scaling. More specifically, we have a situation of bifractality: as discussed above, there are two types of singularities that control all moments, and, as a consequence, the spectrum $\Delta_q$ is formed by two straight lines. 

It should be mentioned that in our analysis we have neglected logarithmic corrections to scaling that are known to arise in the Fisher-KPP model and in further related problems of front propagation. While these corrections do not affect the asymptotic ($N\to\infty$) values of fractal exponents  $\Delta_q$, they are quite substantial if one extracts $\Delta_q$ from the scaling of wave function moments by numerically diagonalizing systems of finite size $N$. The point is that these finite-size corrections to $\Delta_q$ decay with increasing $N$ very slowly, only as $1/\ln N$, thus remaining quite sizeable for Cayley-tree systems with largest $N$ that are still amenable to exact numerical diagonalization. We will return to this issue in Sec.~\ref{s3} in course of the analysis of our numerical data. 

\subsection{Statistics of eigenfunctions vs statistics of LDOS}
\label{s2c}

It is worth pointing out that the exponents $\Delta_q$, Eq.~(\ref{qres}), characterizing the fractality of eigenstates in the intermediate phase on the Cayley tree do not satisfy the symmetry relation 
\be
\label{symmetry}
\Delta_q = \Delta_{1-q}
\ee
that is an exact property of multifractal spectra at Anderson-transition critical points in conventional systems \cite{mirlin06}. Since this fact is a manifestation of an important difference between the peculiar fractal phase on the Cayley tree and conventional critical systems at Anderson-transition points, we discuss it now in some detail. 

An obvious question to be asked is:  How does the system manage to violate the symmetry relation (\ref{symmetry}) if it is exact? To answer this question,  we remind the reader the origin of the relation (\ref{symmetry}). 

Let us denote by $\sigma$ and $\rho$  the real and imaginary parts of the Green functions (\ref{Green}),
\be
\rho = - \frac{{\rm Im}\:G_R}{\pi \nu}, \qquad \sigma  = \frac{{\rm Re}\:G_R - \langle {\rm Re}\:G_R \rangle} {\pi \nu},
\label{rho-sigma}
\ee
where $\nu = - \langle {\rm Im}\:G_R \rangle/\pi$ is the average LDOS. 
Clearly, $\rho$ is the (normalized) fluctuating LDOS. The normalization and the shift of the real part in Eq.~(\ref{rho-sigma}) are chosen in such a way that $\langle\rho\rangle=1$ and $\langle\sigma\rangle = 0$.  The joint distribution function of $\sigma$ and $\rho$ can be expressed through the order-parameter function $Y(Q) = Y(\lambda_1,\lambda_2)$ in the following way\cite{mf94}:
\be
{\cal P}(\sigma,\rho)=\frac{1}{\rho^2}\frac{\partial}{\partial\lambda_1}
(\lambda_1^2-1)\frac{\partial}{\partial\lambda_1}\tilde{Y}(\lambda_1)
\left|_{\lambda_1=\frac{\sigma^2+\rho^2+1}{2\rho}}\right. ,
\label{distr-sigma-rho}
\ee
where
\be
\tilde{Y}(\lambda_1)=\frac{1}{4\pi}\int_{-1}^1\frac{d\lambda_2}{\lambda_1-\lambda_2}
Y({\lambda_1,\lambda_2}).
\label{Y0}
\ee
Integration of Eq.~(\ref{distr-sigma-rho}) over $\sigma$ yields the LDOS distribution \cite{mf94,mirlin00}
\be
{\cal P}(\rho)=\frac{\partial^2}{\partial \rho^2}\int_{\frac{\rho^2+1}{2\rho}}^{\infty}
d\lambda_1 \tilde{Y}(\lambda_1)
\left(\frac{2\rho}{\lambda_1-\frac{\rho^2+1}{2\rho}}\right)^{1/2}\ .
\label{distr-rho}
\ee

The distribution function (\ref{distr-rho}) satisfies a symmetry relation 
\be
\label{symmetry-ldos-distr}
{\cal P}(\rho^{-1}) = \rho^3 {\cal P}(\rho),
\ee
which is exact for a sigma model independently of the spatial geometry and coupling strength.
An equivalent statement is the relation in terms of LDOS moments,
\be
\label{symmetry-ldos-moments}
\langle\rho^q\rangle = \langle\rho^{1-q}\rangle.
\ee
To employ these relations for extracting scaling at Anderson transition, one should open the critical system in a certain way, thus broadening energy levels. (For the closed system with sharp levels the LDOS is a sum of delta functions in energy, and its moments are not defined.) One possibility to do this is to consider a $d$-dimensional critical system of size $L$ coupled at the boundary to metallic electrode and to study physics in the middle of the system. Another possibility is to attribute  to all energy levels an identical width $\eta$ of the order of the mean level spacing $\sim 1/N$, where $N \sim L^d$ is the system volume. It turns out that in any of these cases the resulting scaling of LDOS moments $\langle\rho^q\rangle$ at the Anderson transition will be the same as the scaling of moments of an eigenfunction intensity, $\langle (N|\psi^2|)^{q}\rangle$.  In view of the symmetry of the LDOS moments, Eq.~(\ref{symmetry-ldos-moments}), this implies the symmetry relation (\ref{symmetry}) for the exponents $\Delta_q$ characterizing the statistics of wave functions of a closed system. 

The key point in the above argument is thus a connection between the moments of wave functions of a closed system and moments of LDOS of an open system. Let us explore how it gets violated in the present problem.

Consider first the situation when all levels are broadened with an equal small width $\eta$, which yields exactly the action (\ref{action}). The LDOS distribution function at the root can be expressed through the corresponding order-parameter function $Y(Q) = Y(\lambda_1,\lambda_2)$, Eq.~(\ref{Y}). Substituting Eqs.~(\ref{asymptotic}), (\ref{Phi}), and (\ref{psires}) in Eq.~(\ref{Y}) for  the order-parameter function, and the result in Eq.~(\ref{Y0}), we find the following form of the function $\tilde{Y}(\lambda_1)$ that determines the statistics of local Green functions at the root,
\be
\label{Y0-cayley}
\tilde{Y}(\lambda_1) \simeq \frac {1}{2\pi\lambda_1} \times \left\{
\begin{array}{ll}
1 - \frac{2\pi \eta\lambda_1 N}{\delta_0}\ , & \ 1< \lambda_1 < \lambda_- \:; \\[0.2cm]
1 - c''\left[\frac{2\pi \eta\lambda_1 N^{\alpha_*}}{\delta_0}\right]^{\beta_*}\!, & \ \lambda_-< \lambda_1 < \lambda_+ \:;\\[0.2cm]
0, & \ \lambda_1 > \lambda_+ \ ,
\end{array}
\right.
\ee
where
\be
\label{lambdas}
\lambda_- = \frac{\delta_0}{2\pi\eta} N^{-\frac{1-\alpha_*\beta_*}{1-\beta_*}}, \qquad 
\lambda_+ = \frac{\delta_0}{2\pi\eta} N^{-\alpha_*}.
\ee

Analyzing the joint distribution ${\cal P}(\sigma,\rho)$, we focus on the range of $\rho > 1$. The most interesting part of the distribution is the one corresponding to the fractal behavior  of $\tilde{Y}(\lambda_1)$ represented by the second line Eq.~(\ref{Y0-cayley}). This corresponds to the region   $\lambda_- < (\sigma^2+\rho^2)^{1/2} < \lambda_+$.   Using Eq.~(\ref{distr-sigma-rho}), we get the following behavior of the joint ditribution function in this region
\begin{widetext}
\be
\label{sigma-rho-distr}
{\cal P}(\sigma,\rho) \sim \lambda_+^{-\beta_*} \times \left\{
\begin{array}{ll}
 \rho^{-3+\beta_*}, & \ \ \rho > \sigma\ \  {\rm and}\ \ \lambda_- < \rho < \lambda_+\:; \\
  \rho^{-1-\beta_*} \sigma^{-2+2\beta_*}, & \ \ \rho < \sigma\  \ {\rm and}\ \ (\rho\lambda_-)^{1/2} < \sigma < (\rho\lambda_+)^{1/2} \:.
  \end{array}
  \right.
\ee
\end{widetext}
We thus see how the exponents $\beta_*$ and $\alpha_*$ show up in the non-trivial power-law behavior of the distribution function and in the borders of this behavior. For $(\sigma^2 + \rho^2)^{1/2} > \lambda_+$ the distribution function is strongly (exponentially) suppressed. 

We turn now to the LDOS distribution function ${\cal P}(\rho)$.   Substituting Eq.~(\ref{Y0-cayley}) into Eq.~(\ref{distr-rho}), we get 
\be
\label{ldos-distr}
{\cal P}(\rho) \sim \lambda_+^{-1/2}\: \rho^{-3/2}, \ \ \ \ \ \lambda_+^{-1}  \lesssim \rho \lesssim \lambda_+ \:.
\ee
Outside of the region $\lambda_+^{-1}  \lesssim \rho \lesssim \lambda_+$ the distribution decays fast. 
Interestingly, the power-law scaling (\ref{ldos-distr})
 of the LDOS distribution is characterized by the exponent 3/2, independently of the value of $\beta_*$.  The fractality of the system enters however  Eq.~(\ref{ldos-distr}) via the borders of the power-law regime. 
The LDOS distribution (\ref{ldos-distr}) can be also obtained from the joint distribution (\ref{sigma-rho-distr}) by integrating it over $\sigma$. (The $\sigma$ integral is given by the upper limit, $\sigma\sim (\rho\lambda_+)^{1/2}$ in view of $\beta_* > 1/2$.) 
 
 The distribution (\ref{ldos-distr}) implies the following scaling of moments:
\be
\label{ldos-moments}
\langle\rho^q\rangle \sim \left \{
\begin{array}{ll}
\lambda_+^{-q}, & \ \ \ q < 1/2, \\[0.2cm]
\lambda_+^{q-1}, & \ \ \ q >1/2.
\end{array} 
\right.
\ee
Clearly, Eqs.~(\ref{ldos-distr}) and (\ref{ldos-moments}) satisfy the relations (\ref{symmetry-ldos-distr}) and (\ref{symmetry-ldos-moments}). 
We can now compare this behavior of LDOS moments with that of wave function moments, Eq.~(\ref{qres}). Choosing the level broadening $\eta$ of the order of the level spacing in the whole system, $\delta_N = \delta_0/N$, we observe that the scaling of $\langle\rho^q\rangle$ and $\langle(Nu)^q\rangle$ at $q>1/2$ is the same, see Eq. (\ref{lambdas}).  On the other hand, at $q<1/2$ the scaling is different: the moments of LDOS respect the symmetry (as they should), while the moments of wave functions scale differently. 

We offer the following physical explanation of this difference in the behavior of the moments $\langle\rho^q\rangle$ and $\langle(Nu)^q\rangle$ at $q < 1/2$. These moments are determined by probabilities of atypically small values of the corresponding variables. The LDOS get contribution not from a single level but from many levels around given energy. The probability to have an anomalously small LDOS will scale in the same way as a probability to have an anomalously small wave function intensity only if wave functions at nearby energies are fully correlated. Such strong correlations are indeed an important property of the Anderson transition\cite{mirlin00,evers08}. We conclude that the correlations between different wave functions behave in an essentially different way (are much weaker for close energies) in the intermediate fractal phase on Cayley tree.  

Another way to broaden the levels is to open the system at the boundary (i.e., to bring all boundary sites in a contact with a perfect metal). As mentioned above, for a conventional critical (Anderson-transition) system, this would lead (far from the boundary) to essentially the same result as broadening all levels by $\eta \sim \delta_N$. The situation is again qualitatively different for the present problem. Opening the system at the boundary means supplementing the recurrence relation (\ref{recurse}) with a boundary condition $\Psi_0(\lambda_1,\lambda_2)$ having a form of the kink that goes to zero at a characteristic scale that does not depend on the system size, $\lambda_1 \sim 1$. In this situation, iteration of Eq.~(\ref{recurse}) will converge to a solution of the self-consistency equation. The resulting LDOS distribution ${\cal P}(\rho)$  and LDOS moments  $\langle\rho^q\rangle$ will show at large $N$ an $N$-independent behavior characteristic for conventional (ergodic) delocalized systems. Thus, contrary to Anderson-transition critical points, the fractality of LDOS in the intermediate phase on a Cayley tree disappears when one opens the system at the boundary.

Finally, one can consider a situation in which all levels are broadened with an equal width $\eta$, and an order of limits opposite to the one appropriate for extracting the statictics of an eigenfunction is considered, i.e., $N\to \infty$ at fixed small $\eta$. For a conventional system at Anderson transition, in this situation one will find a scaling of LDOS moments $\langle\rho^q\rangle \sim L_\eta^{-\Delta_q}$, where $L_\eta$ is a characteristic length set by $\eta$.  Thus, one can probe the (multi-)frac\-ta\-lity at Anderson transition also in this way. On the other hand, for the present problem, this order of limits will eliminate the fractality. Indeed, also in this case, the iteration of Eq.~(\ref{recurse}) will converge to a solution of the self-consistency equation which, in the delocalized phase, does not depend on $\eta$ for small $\eta$. Therefore, the LDOS distribution and moments will be essentially independent of $\eta$, thus showing no trace of fractality characteristic for individual eigenfunctions. 

\subsection{Anderson model  with $n=1$}

As we have demonstrated above, the delocalized phase of the unitary-class sigma model on a finite Cayley tree  is subdivided
into ``ergodic'' and ``non-ergodic'' (fractal)  phases as judged by the statistics of wavefunctions at the root. The obtained results on the phase boundary and the fractal exponents are determined by the spectrum $\epsilon_\beta$, Eq.~(\ref{specUnitary}), of the linearised recursive equation describing the integration over successive layers of the tree. This analysis is rather general in the sense that it can be applied also to other models of localization on Cayley trees. The key point is that the spectrum $\epsilon_\beta$, although somewhat different for different models, has exactly the same qualitative properties as listed below Eq.~(\ref{specUnitary}). 

In particular, our analysis can be straightforwardly applied to sigma models of other symmetry classes (orthogonal and symplectic), corresponding to systems with preserved time-reversal invariance. The explicit form of the kernel $L(t)$ and of the corresponding eigenvalues $\epsilon_\beta$ for these models can be found Ref.~\onlinecite{efetov87,verbaarschot88}. All results are qualitatively the same as in the unitary-symmetry model. Moreover, curves for different sigma models merge in the large-$m$ limit, as discussed below. 

Our analysis can be also extended to the original Anderson model with $n=1$ orbital per site [Eq.~(\ref{H}) where we set, following standard convention,  $t=1$]. In this model, the recursion relation that is a counterpart of  Eq.~(\ref{recsimple}) involves a function of two variables, as it is connected with the joint probability distribution function of real and imaginary parts of the Green function, see Refs.~\onlinecite{abouchacra73,mirlin91}. As a result, the role of $\epsilon_\beta$ is played by the largest eigenvalue of a certain $\beta$-dependent linear  integral operator. While the eigenvalues $\epsilon_\beta$ depend now on the specific distribution of disorder and cannot be obtained in a closed analytical form, their gross features are the same as for the sigma model [see text below Eq.~(\ref{specUnitary}) and Fig.~\ref{eigen}]. Thus, on the qualitative level, all the conclusions obtained above for the sigma model remain applicable also for the $n=1$ Anderson model. 
To obtain explicit analytical results for the $n=1$ model, we consider the limit of large connectivity, $m\gg 1$. In this case, the relevant values of disorder are $W\gg 1$, which allows us to use the large-$W$ approximation for the eigenvalues $\epsilon_\beta$. Such an approximation was developed in Ref.~\onlinecite{abouchacra73} for the eigenvalue $\epsilon_{1/2}$ determining the position of the Anderson transition, with the result
\be
\label{abou-transition}
\epsilon_{1/2} \simeq \frac{4}{W} \ln\frac{W}{2}.
\ee
It is not difficult to generalize the corresponding derivation in Ref.~\onlinecite{abouchacra73} onto the case of $\beta \ne 1/2$. This yields
\be
\label{abou}
\epsilon_{\beta}\simeq \frac{1}{\beta-1/2}\frac{1}{W-4/W}\left[\left(W/2\right)^{2\beta-1}-\left(W/2\right)^{-2\beta+1}\right].
\ee
 The approximation (\ref{abou}) preserves all properties of $\epsilon_\beta$, including the exact symmetry $\beta\to 1-\beta$ and the exact identity $\epsilon_1 = 1$.  Using this expression [which is a counterpart of the sigma model formula (\ref{specUnitary})], we can evaluate the exponents $\beta_*$ and $\alpha_*$ that determine, according to Eqs.~(\ref{presult}) and (\ref{qres}), the wave function statistics.  
Already for $m=2$ this gives quite a decent approximation which is, however, not controlled parametrically. It becomes controlled in the large-$m$ limit that we are going to discuss now.

\subsection{Large connectivity: Anderson model and sigma model}

\begin{figure}
\centering
\includegraphics[width=.5\textwidth]{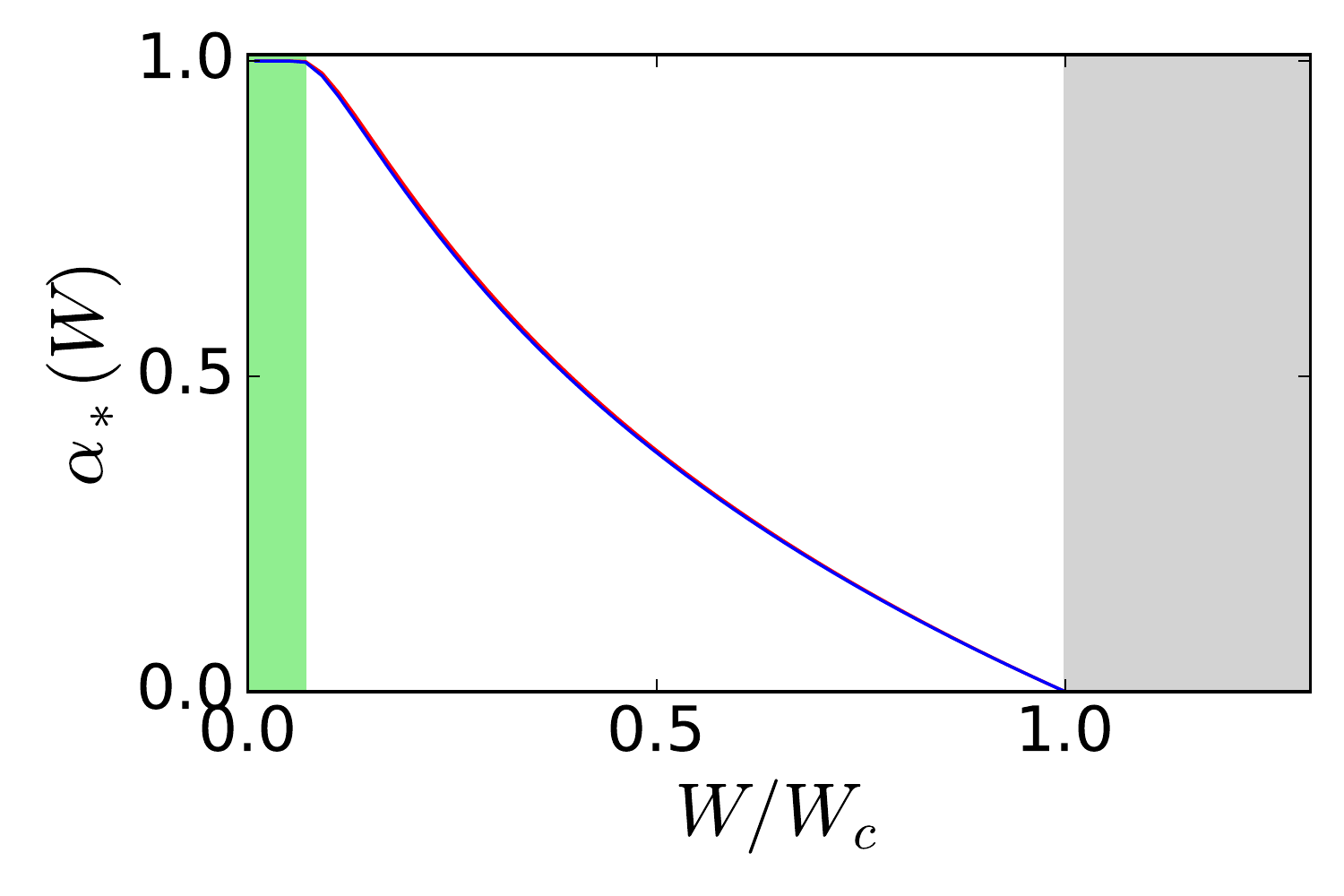}
\caption{Fractal exponent $\alpha_*$ controlling the scaling of the second moment $P_2$ of the wave function intensity (and also of all moments $P_q$ with $q>q_*$)   for the connectivity $m=16$ as a function of relative strength of disorder $W/W_c$.  Red line: unitary sigma model (corresponding to the Anderson model with many orbitals per site, $n\gg 1$); blue line: Anderson model with $n=1$ for which the large-$m$ approximation (\ref{abou}) was used. (Two lines are almost indistinguishable.) The sigma-model coupling is connected to the Anderson-model disorder via $g/g_c = (W_c/W)^2$. 
The critical points for these models are $W_c=326$ and $g_c=2.2\cdot 10^{-4}$, respectively.
As in Fig.~\ref{figlambda}, three phases are clearly seen: localized (gray) with $\alpha_* =0$, delocalized fractal with $0 < \alpha_* < 1$, and delocalized ergodic with $\alpha_*=1$ (green). }
\label{largem}
\end{figure}

As was pointed out in Ref.~\onlinecite{mirlin97}, for large connectivity of a Bethe lattice one expects a quantitative equivalence between the $n=1$ Anderson model and the sigma models representing its large-$n$ limit. To verify this prediction for the present problem, we plot in Fig.~\ref{largem} the fractal exponent $\alpha_*$ for the unitary sigma model and the $n=1$ Anderson model on a Cayley tree with $m=16$ as a function of disorder (normalized to its critical value) $W/W_c$. For the Anderson model, we have used the large-$W$ approximation (\ref{abou}). For the sigma-model, we have used the expression of the sigma-model coupling $g$ in terms of the microscopic model, $g= n^2/W^2$, yielding the identification $g/g_c = (W_c/W)^2$. Remarkably, already for moderately large $m$, such as $m=16$, the disorder dependences of the fractal exponent in the Anderson and sigma models become essentially indistinguishable, see Fig.~\ref{largem}. 

In fact, the disorder dependence of the fractal exponents takes a very simple analytic form in the asymptotic limit $m\to\infty$. 
Specifically, using  Eq.~(\ref{abou}) with $W\gg 1$ [or Eq.~(\ref{specUnitary}) with $g \ll 1$] for the eigenvalue $\epsilon_\beta$, we find 
from Eqs.~(\ref{lambda-opt1}), (\ref{alphadef}) a logarithmically slow variation of the exponents $\beta_*$ and $\alpha_*$ in the non-ergodic delocalized phase, $W_e \ll W \ll W_c$:
\bea
\beta_*^{-1} &=& 2 - \frac{2}{2\ln W - \ln m};\\[0.2cm]
\alpha_* &=& 2 - \frac{2 \ln W}{ \ln m}.
\eea
The boundaries of this regime are given on the logarithmic scale by
\bea
\ln W_e  \simeq \frac{1}{2}\ln m; 
\label{We-large-m}\\
\ln W_c \simeq \ln m.
\label{Wc-large-m}
\eea
(Here we neglected subleading corrections of the type $\ln\ln m$ and const.) As follows from Eqs.~(\ref{We-large-m}) and (\ref{Wc-large-m}), for a Cayley tree with large $m$ the non-ergodic domain occupies, on the linear scale of disorder, the dominant part of the delocalized phase. This is well illustrated by the case $m=16$ in Fig.~\ref{largem}. 

\section{Wavefunction statistics from exact diagonalization: Cayley tree vs random regular graphs}
\label{s3}

In the preceding Section, we have studied analytically the statistics of wave functions at a root of a finite Cayley tree. We have found that, in a large part of the delocalized phase, the eigenfunctions are fractal and have determined the corresponding fractal exponents.  These results differ crucially from that obtained analytically\cite{sparse} and numerically\cite{tms16} for the RRG model where the eigenstates are ergodic in the whole delocalized phase. 
In this section, we verify our analytical predictions by performing a detailed numerical analysis of the eigenfunction statistics in the delocalized phase on Cayley tree with $m=2$.  We will compare and contrast these results with those on RRG.  To make this comparison particularly transparent, we have performed the simulation for the RRG model fully analogous to that in Ref.~\onlinecite{tms16} but using the RRG of exactly the same sizes $N$ as Cayley trees studied here.

\subsection{Numerical results}

\begin{figure*}[htp]
\centering
\includegraphics[width=.4\textwidth]{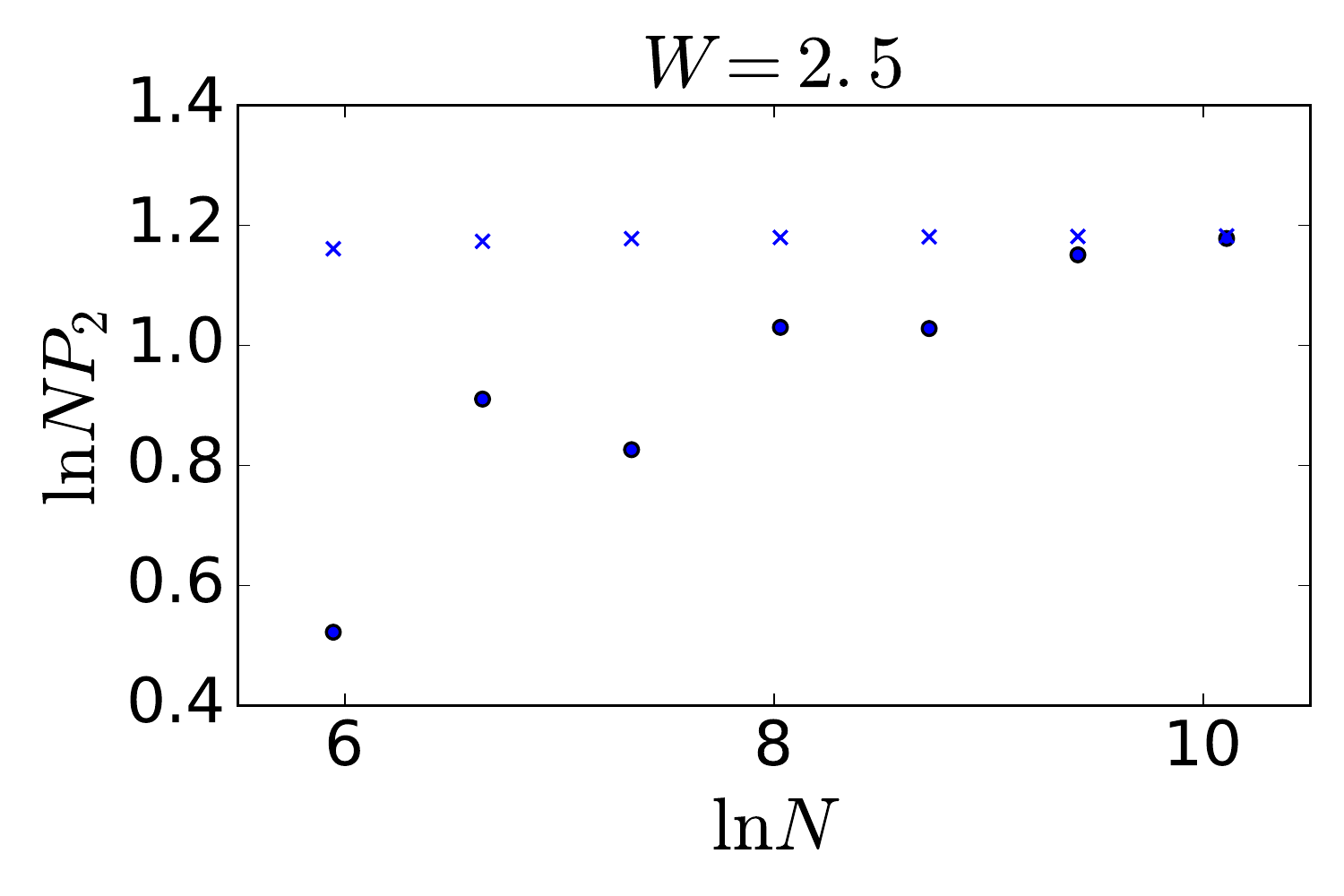}\hspace{0.05\textwidth}
\includegraphics[width=.4\textwidth]{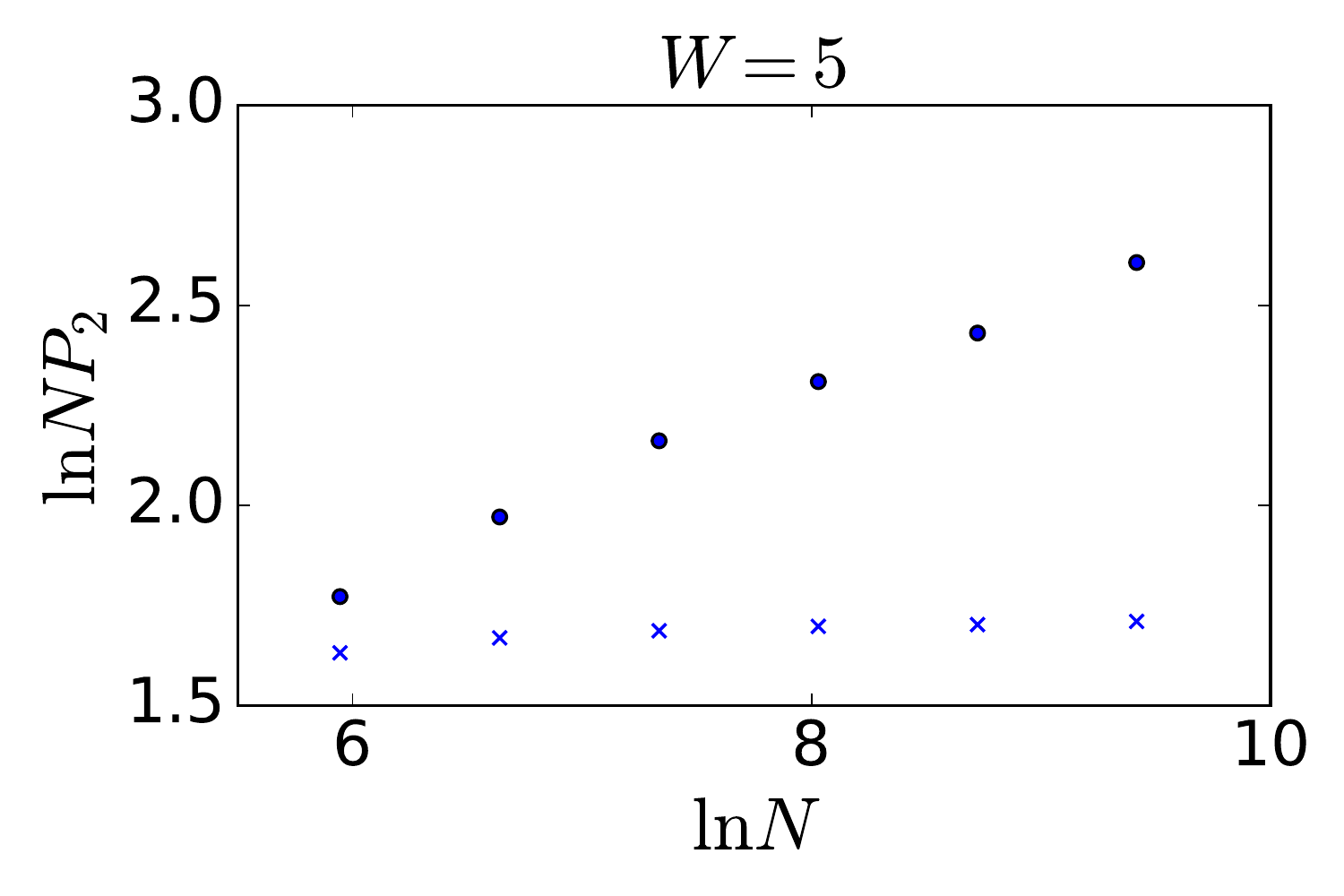}
\\
\includegraphics[width=.4\textwidth]{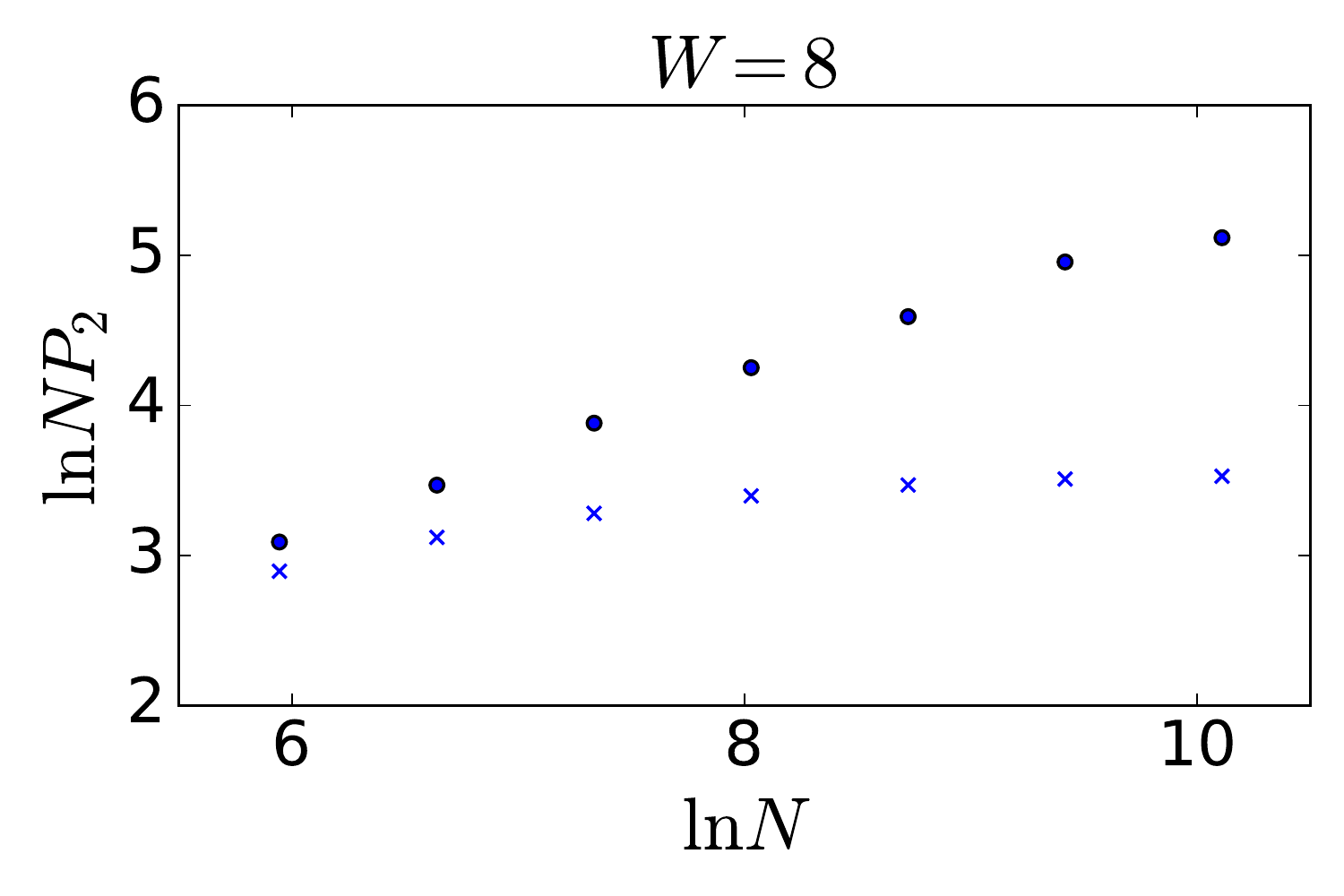}\hspace{0.05\textwidth}
\includegraphics[width=.4\textwidth]{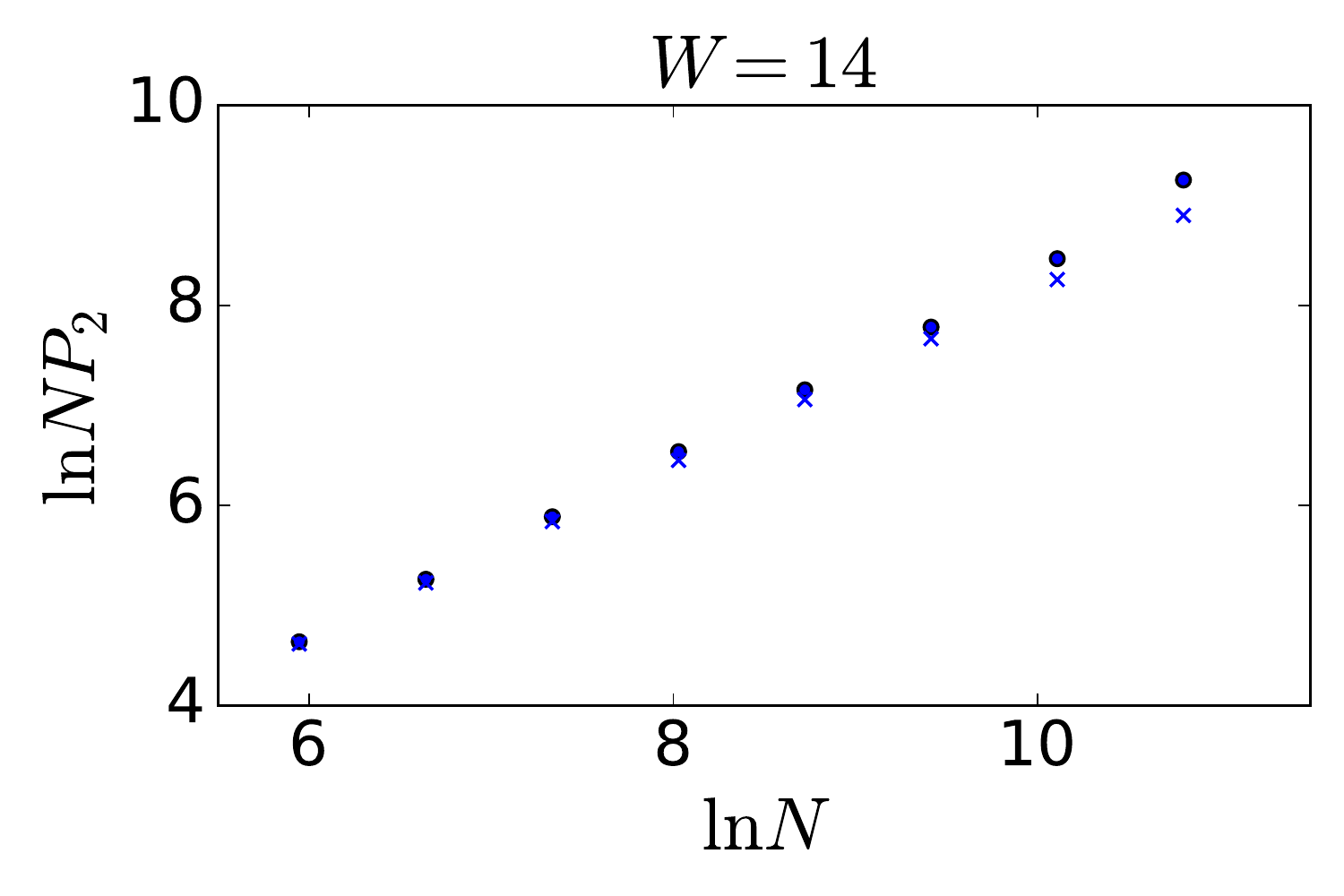}
\caption{System-size dependence of the second moment of wavefunctions, $NP_2$, at the root of a Cayley tree ($n=1$ Anderson model, connectivity $m=2$; circles) and on RRG (crosses) for disorder $W=2.5, 5, 8,$ and 14.  For the Cayley tree model, the weakest disorder corresponds to the ergodic phase $W < W_e$, the other three to the intermediate delocalized fractal phase ($W_e < W < W_c$).  For the RRG model, the whole delocalized phase $W < W_c$ is ergodic. }
\label{fig:tvg}
\end{figure*}

We start with the comparison of the behavior of the second moment $P_2$ on Cayley tree and on RRG. 
In Fig.~\ref{fig:tvg} we show the dependence of  $NP_2$ on the system size for several representative values of disorder. 
We choose to plot $NP_2$ since it should saturate in the limit $N\to \infty$ in the conventional delocalized phase.
The numerical results fully support our analytical fundings, demonstrating clearly two key features: (i) emergence of intermediate ``non-ergodic'' delocalized phase on the Cayley tree, and (ii) dramatic difference between the fractal behavior on Cayley tree in this intermediate phase and the conventional (``ergodic'') behavior in the RRG model.  

Let us briefly comment on each of the panels of Fig.~\ref{fig:tvg}. For weak disorder ($W=2.5$) we observe that $NP_2$ saturates in both Cayley tree and RRG models at approximately the same value which is only slightly higher than that for the Gaussian orthogonal ensemble (GOE), $NP_2 = 3$.  For RRG this saturation has been shown numerically in Ref.~\onlinecite{tms16}. For Cayley tree this behavior demonstrates that the point $W=2.5$ belongs to the ergodic part of the delocalized phase, $W < W_e$, see Sec.~\ref{s2}.  It is worth mentioning that oscillations of $NP_2$ on Cayley tree at relatively small system sizes are not statistical fluctuations but rather even-odd finite-size oscillations that are remnants of clean Cayley tree. Another indication of the ``almost clean'' character of the system at these small sizes is the fact that $NP_2$ is substantially smaller than its GOE value. At the largest studied system sizes (12 to 14 generations), the oscillations get strongly damped and $NP_2$ saturates. 

The behavior on Cayley tree becomes essentially different for further three values of disorder, $W= 5, 8$, and 14, shown in Fig.~\ref{fig:tvg}. Here $NP_2$ grows as a power law of $N$: the Cayley-tree model is in the intermediate fractal delocalized phase, $W_e < W < W_c$, see Sec.~\ref{s2}. 
This is in stark contrast with the saturation of $NP_2$ (which is a manifestation of the conventional ``ergodic'' behavior) in the RRG model clearly observed for $W=5$ and $W=8$, in full agreement with our earlier results\cite{tms16}. Thus, the $W=5$ and $W=8$ panels of Fig.~\ref{fig:tvg} visualize in a particularly clear way the key difference between the Cayley-tree and the RRG models: the existence of an intermediate fractal delocalized phase in the former and the ergodicity of delocalized states in the latter.

For $W=14$, which is quite close to the critical disorder $W_c \simeq 17.5$,  this saturation in the RRG model cannot be achieved with the studied system sizes. The reason for this was explained in detail in Ref.~\onlinecite{tms16}: the RRG  ``correlation volume'' $N_c(W)$ increases exponentially near the transiton, becoming comparable to our largest system size at $W=14$. For sizes $N$ smaller than $N_c$ the RRG model is essentially in the critical state, implying ``almost localized'' behavior of eigenfunctions. This explains a similarity of the behavior of $NP_2$ on RRG to that in the root of Cayley tree where eigenfunctions show fractal behavior with an exponent $\alpha_*$ close to zero (which is its ``localized'' value). Only for our largest system sizes the $W=14$ data on RRG start to deviate downwards, thus showing a trend to saturation.  (To observe full development of this saturation for RRG with $W=14$, system sizes would be needed that are beyond our computational capabilities.)

\begin{figure}
\centering
\includegraphics[width=0.45\textwidth]{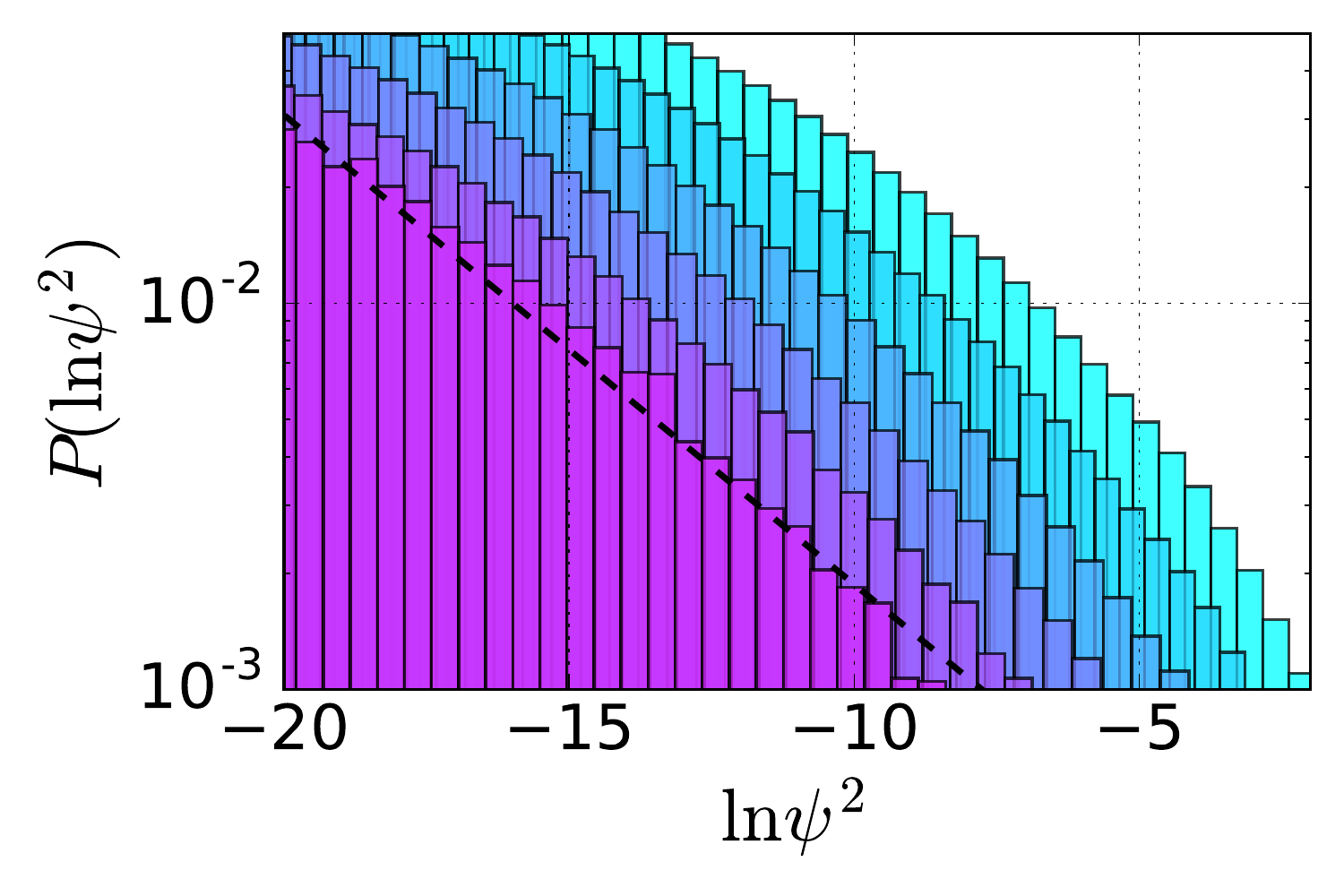}
\caption{Distribution function of $\ln u \equiv \ln |\psi|^2$ for $W=14$ at a root of a Cayley tree with the number of generations $s_0$ from 9 to 14. A power-law distribution ${\cal} P(u) \propto u^{\beta_*-2}$ corresponds to a straight line in this plot. It is seen that a range of power-law behavior develops with increasing number of generations and moves to the left with a constant speed, in agreement with the analytical result (\ref{presult}). The dashed line corresponds to the value $\beta_* = 0.72$ extracted from the behavior of moments, Fig.~\ref{fig:bi}.}
\label{fig:pdf}
\end{figure} 

After having discussed the second moment $P_2$, we turn to numerical analysis of the whole distribution function of the eigenfunction intensity, ${\cal P}(u)$. In the preceding Section, we have found analytically that the dominant part of this distribution function acquires a power-law form, Eq.~(\ref{presult}). It is convenient to transform Eq.~(\ref{presult}) to a distribution of the logarithm $\ln u$, which yields
\be
\label{Plnu}
\ln\mathcal{P}(\ln u)=(\beta_*-1)\ln u + (-1+\alpha_*\beta_*)\ln N
\ee
with a support at 
\be
\label{Plnu-support}
-\frac{1-\alpha_*\beta_*}{1-\beta_*}\ln N\lesssim \ln u\lesssim -\alpha_*\ln N.
\ee
Thus, we expect a linear part on the plot of $\ln\mathcal{P}(\ln u)$ with a slope $\beta_*-1$ that becomes increasingly more developed with increasing system size $N$ and moves to the left with a constant speed with respect to $\ln N$. To verify these predictions, we plot in Fig.~\ref{fig:pdf} the distribution $\ln\mathcal{P}(\ln u)$ for $W=14$ for $m=2$ Cayley trees of different sizes (number of generations $s_0$ from 9 to 14). We observe that a region of linear slope indeed develops with increasing system size and moves to the left with a constant speed with respect to $s_0$ (or, equivalently, with respect to $\ln N$), in agreement with analytical results, Eqs.~(\ref{Plnu}) and (\ref{Plnu-support}).

\begin{figure}
\centering
\includegraphics[width=0.45\textwidth]{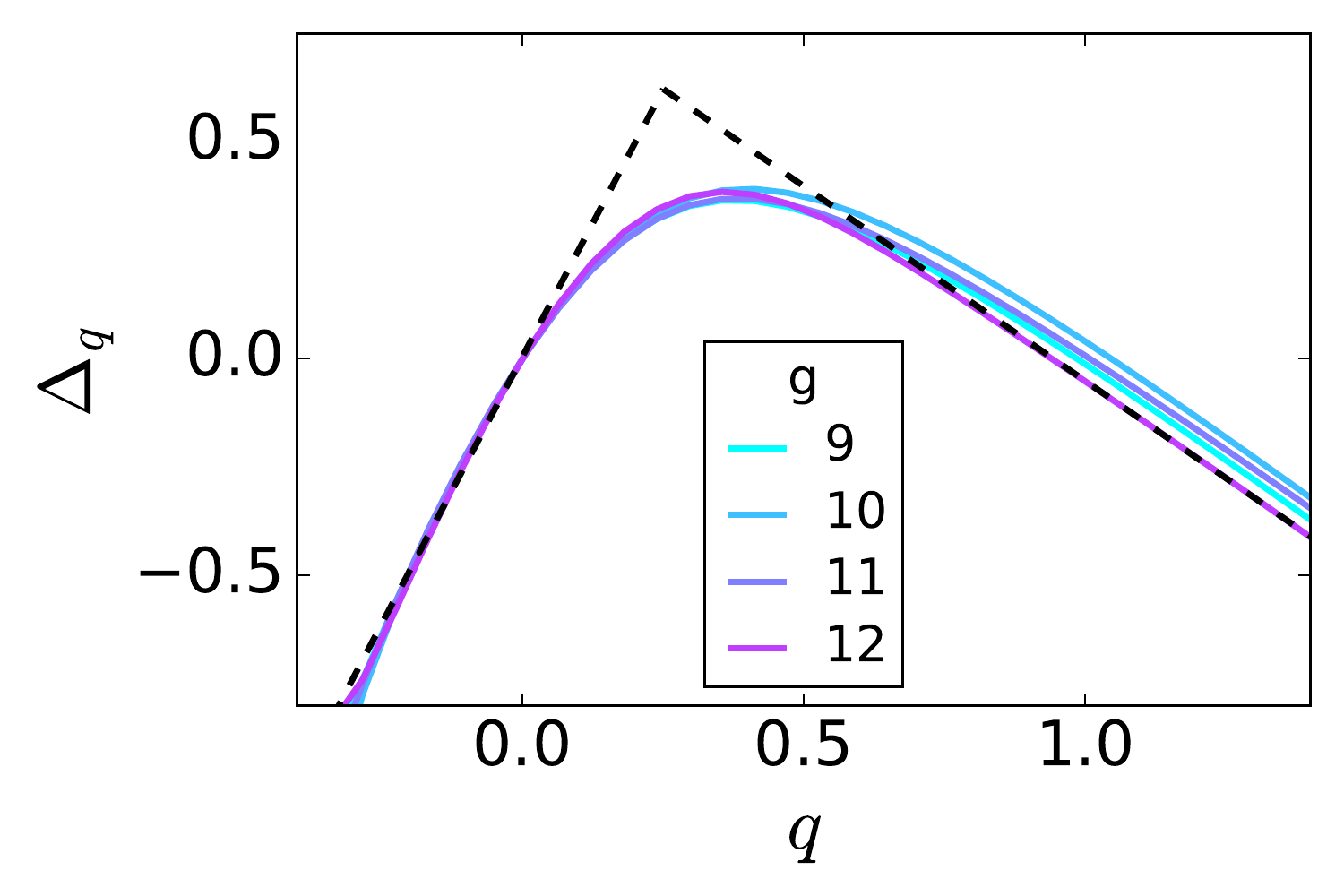}
\caption{Anomalous dimensions $\Delta_q$ characterizing fractality of eigenfunction intensity at the root of a Cayley tree with disorder $W=14$, as determined for systems of different number of generations $s_0$ (from 9 to 12).  Dashed lines represent a fit to Eq.~(\ref{qres}) with $\beta_* \simeq 0.72$ and $\alpha_* \simeq 0.08$.}
\label{fig:bi}
\end{figure} 

As a further characterization of our numerical data, we study the anomalous fractal dimension $\Delta_q$ as function of $q$.
Our analytical result, Eq.~(\ref{qres}), predicts a bifractal behavior. To verify this, we plot in Fig \ref{fig:bi} numerically extracted values of $\Delta_q$ with $q$ between $-0.25$ and 1.5 for $W=14$ and for systems for the number of generations from 9 to 12. The results support the bifractal behavior (\ref{qres}).  As expected, the cusp in $\Delta_q$  at $q=q_*$ is rounded due to finite-size effects. The convergence for $q$ in a close vicinity of $q_*$ appears to be quite slow, and considerably larger system sizes would be needed to see it more clearly. Since the slopes of $\Delta_q$ at $q>q_*$ and $q<q_*$ are expressed in terms of $\beta_*$ and $\alpha_*$ via Eq.~(\ref{qres}), we can extract the values of these exponents from the numerically found slopes. This yields  $\beta_* \simeq 0.72$ and $\alpha_* \simeq 0.08$. Using this value of $\beta_*$, we have plotted a dashed line in Fig.~\ref{fig:pdf}.
It is seen that this value of $\beta_*$ indeed describes correctly the power-law distribution of wave function intensities, Eq.~(\ref{Plnu}).

\begin{figure}
\centering
\includegraphics[width=.5\textwidth]{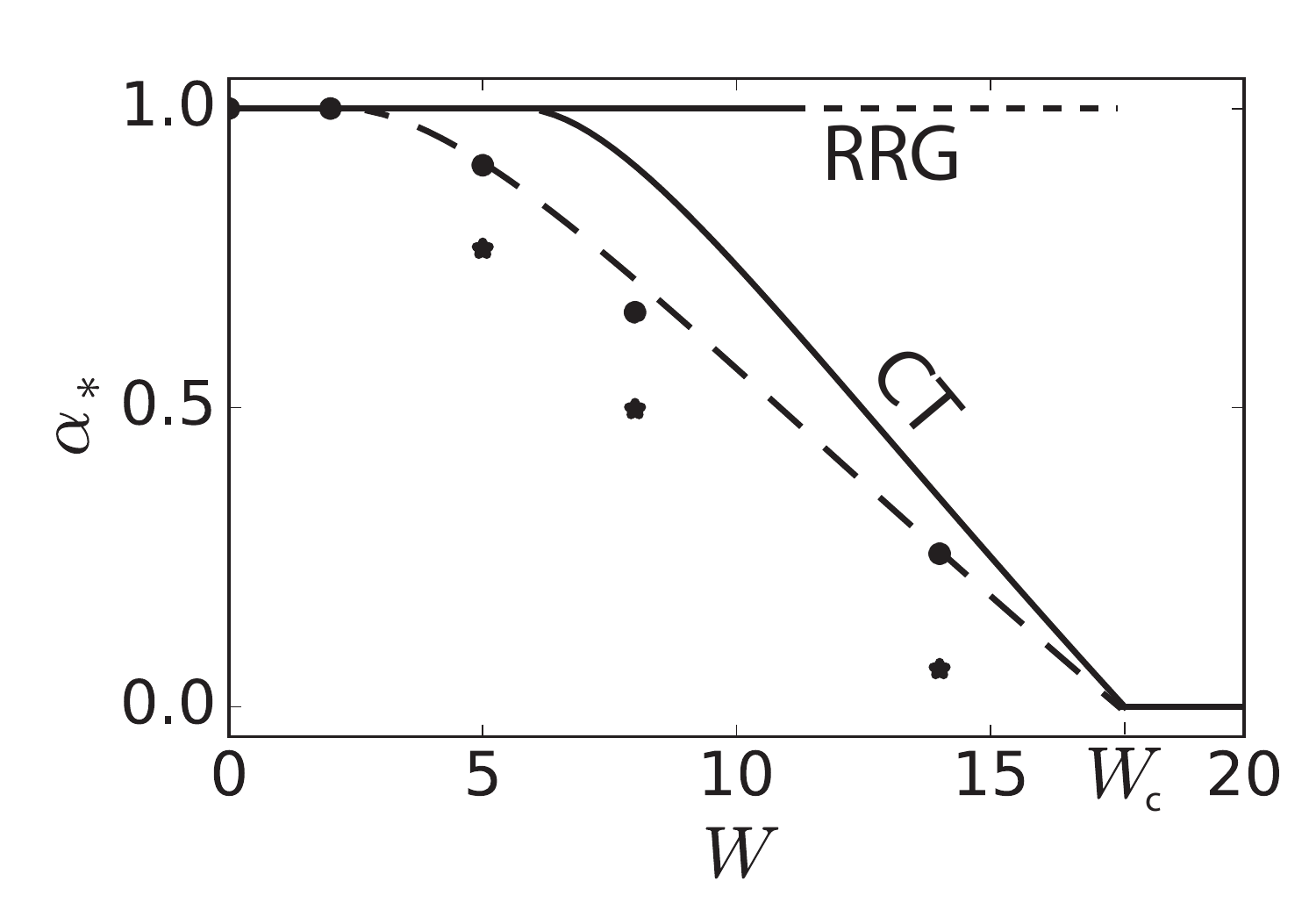}
\caption{Disorder dependence of the fractal exponents $\alpha_*$ (defined in the limit $N\to\infty$) in the delocalized phase of $n=1$ Anderson model with $m=2$. CT:  Cayley tree; full line: analytical result, see Sec.~\ref{s2}, in combination with the large-$W$ approximation (\ref{abou});
stars: numerically extracted $\alpha_*^{(N)}$; dots: asymptotic ($N\to\infty$) value $\alpha_*$ obtained from $\alpha_*^{(N)}$ by taking into account the finite-site correction (\ref{alpha-N}); dashed line: fit to finite-size-corrected numerical data.  
RRG: random regular graphs; solid  line: numerically extracted $\alpha_*$  for RRG (Ref.~\onlinecite{tms16} and this work); dashed line: extrapolation on the basis of analytical prediction\cite{sparse} supported by numerical data of Ref.~\onlinecite{tms16} and of this work (full saturation of IPR is not achievable for this disorder values in view of system-size limitations).  }
\label{alpha}
\end{figure}

In Fig.~\ref{alpha} we show the disorder dependence of the fractal exponent $\alpha_*$ for the $n=1$ Anderson model on a Cayley tree with the connectivity $m=2$, as estimated from our numerical results. Qualitatively, this behavior is the same as in the $n\gg 1$ model (sigma model) on a Cayley tree with $m=2$, Fig.~ \ref{figlambda}, as well as is the large-$m$ models, Fig.~ \ref{largem}. For comparison, we also show by a line the value $\alpha_*=1$ corresponding to ergodic delocalized wavefunctions in the RRG model.

\subsection{Finite-size corrections to scaling on the Cayley tree}
\label{s3.1}

In Fig.~\ref{alpha} we 
compare the data points for the fractal exponent $\alpha_*$ as obtained from exact numerical diagonalization of the Anderson model on the Cayley tree (star symbols) with the analytical formula derived in Sec.~\ref{s2} combined with the large-$W$ approximation (\ref{abou}) for the eigenvalues $\epsilon_\beta$. While two dependences are very similar, we observe a clear downward deviation of numerical data as compared to the analytical curve. Clearly, the large-$W$ formula  (\ref{abou}) is only an approximation for moderate values of $W$ that are of interest for the case of connectivity $m=2$. While this accounts for a part of the deviation, this is is not sufficient to explain it completely. Indeed, it is known that the approximation (\ref{abou}) reproduce with a very good accuracy the critical value $W_c \simeq 17.5$ (which is the point in Fig.~\ref{alpha} where the analytical curve for $\alpha_*(W)$ reaches zero). On the other hand, the numerical data appear to suggest a considerably smaller value $W_c \simeq 15$.  Thus, there should be a reason why our numerical data for $\alpha_*$ are substantially smaller, in the range of not too small $W$, than the actual values of $\alpha_*$. 

A well-known source for deviation of numerical data for exponents from their true values are finite-size effects. Usually finite-site effects in critical phenomena (such as, e.g., Anderson transition) are not well controlled (since their are governed by subleading exponents that are usually not known analytically). However, they usually vanish in a power-law fashion with the system size $N$ and are thus quite small for largest available $N$. As we briefly discuss now, the finite-size effects in the present problem are essentially different in two ways. On one hand, they decay very slowly, only as $1/\ln N$, thus producing a substantial deviation of the exponent even for largest $N$ amenable to numerical diagonalization. On the other hand, the leading correction is known analytically and thus can be taken into account. 

The finite-time correction to the position (and thus to the velocity) of the front was first calculated, for the case of Fisher-KPP equation by Bramson\cite{bramson78}. Later, it was shown that an analogous result applies to a broad class of problems of front propagation\cite{ebert00,brunet15}.  It is thus very plausible that the same form of the correction applies also to the present problem (although a rigorous proof of applicability of this statement remains a prospect for future work). When applied to our problem and translated to our notations, the result of Ref.~\onlinecite{ebert00,brunet15} for the leading-order correction reads 
\be 
\alpha_*^{(N)} = \alpha_* - \frac{3}{2\beta_*\ln N},
\label{alpha-N}
\ee
where
$\alpha_*$ is the true asymptotic ($N\to\infty$) value of the exponent and $\alpha_*^{(N)}$ is its apparent value as obtained for systems of size $N$. Thus, the finite-size correction implies a downward shift of $\alpha$: eigenstates of a finite system appear to be more fractal than they would be in the limit of $N\to\infty$. To estimate the magintude of the correction for realistic $N$, we take $\ln N \simeq 10$ and $\beta_* \simeq 3/4$ (which is in the middle of the full range of $1/2 < \beta_* < 1$ in the fractal phase), which yields the characteristic value $\alpha_* - \alpha_*^{(N)} \simeq 0.2$. This perfectly explains the deviation between the numerical data and the analytical result in the range of intermediate $W$, and thus the apparent shift of $W_c$. 

We have taken into account the finite-size effects according to Eq.~(\ref{alpha-N}); the correspondingly corrected\cite{note-beta} numerical data are shown by dot symbols in Fig.~\ref{alpha}. These corrected data are fitted well by a dependence $\alpha_*(W)$ that has essentially the same shape as the line obtained in high-$W$ approximation, with the same $W_c \simeq 17.5$ but  with somewhat smaller $W_e$, see dashed line in Eq.~(\ref{alpha-N}). This line represent thus our estimate for the asymptotic behavior of the fractal exponent $\alpha_*(W)$. 

The accuracy of our finite-size-corrected result for $\alpha_*(W)$ (dot symbols in Fig.~\ref{alpha}) can be estimated as $\pm 0.1$ on the basis of the leading-order correction and taking into account that higher terms are suppressed by additional powers of $1/(\ln N)^{1/2}$. 
Since one cannot increase substantially $\ln N$ for matrices amenable to exact diagonalization, it would be difficult to improve substantially this accuracy by a direct analysis of statistics of eigenfunctions. On the other hand, one can find the asymptotic fractal exponent $\alpha_*(W)$ of the Cayley-tree problem with a substantially better accuracy by using our analytical results of Sec.~\ref{s2} and studying the front evolution for the non-linear equation corresponding to the $n=1$ Anderson model\cite{abouchacra73,mirlin91}  with a numerical pool method. Some related work was done in Ref.~\onlinecite{abouchacra73,miller94,monthus09}. Very recently, a related population-dynamics algorithm was implemented in Ref.~\onlinecite{altshuler16}; we will comment on this work in more detail in Sec.~\ref{s4}.  It was found in various models of front propagation\cite{brunet97} that the correction to velocity  due to finite size $N_p$ of the pool scales as $1/(\ln N_p)^2$. Since the velocity determines our fractal exponent $\alpha_*$, we expect that it should be possible to obtain  this exponent by the pool method with the accuracy of order $\sim 1\%$. Once $\alpha_*$ and $\beta_*$ are determined numerically, 
one can use our Eq.~(\ref{qres}) to obtain the spectrum of fractal exponents characterizing the eigenfunction statistics at the root of the Cayley tree.

\section{Summary}
\label{s4}

In this paper, we have studied analytically and numerically eigenfunction statistics in a disordered system on a finite Cayley tree. We have shown that the distribution of eigenfunction amplitudes at the root of the tree is fractal (``non-ergodic'') in a large part of the delocalized phase. We have determined fractal exponents characterizing the statistics and the scaling of moments in this peculiar phase. These exponents are expressed,
Eq.~(\ref{qres}), in terms of the decay rate $\beta_*$ and the velocity $v_{\beta_*}=\frac{\alpha_*}{\ln m}$ in a problem of propagation of a front between unstable and stable phases. Our findings imply a crucial difference between a loopless Cayley tree and a locally tree-like structure without a boundary (random regular graph, RRG) where extended eigenfunctions are ergodic. We have also performed numerical simulations of both models (Cayley tree and RRG) that fully support the analytical results. 

We have emphasized the very peculiar character of fractality on the Cayley tree whose existence depends on the order of limits $\eta \to 0$ and $N\to\infty$, where $\eta$ is the level broadening and $N$ the system size. While probing the statistics of individual eigenfunctions, we take the limit $\eta\to 0$ first, which results in a fractal dependence of eigenfunction moments on $N$.  (More generally, this situation is realized if $\eta N^{\alpha_*} \to 0$.) If the opposite order of limits is considered ($N\to\infty$ first, or, more generally, $\eta N^{\alpha_*} \to \infty$), the LDOS moments become $\eta$-independent, contrary to Anderson-transition critical points where they would scale in a fractal way with $\eta$. Similarly, opening the Cayley tree at the boundary (i.e., connecting the boundary to a ``metallic'' system) eliminates the fractal scaling of the LDOS moments with $N$. 

Before closing the paper, we make several comments:

\begin{enumerate}

\item
It is instructive to pinpoint a key distinction between the analytical calculations of eigenfunction moments for locally tree-like graphs without boundary (RRG or SRM, Ref.~\onlinecite{sparse}), on one hand, and Cayley tree (this work), on the other hand, that is responsible for very different behavior (ergodic {\it vs} fractal) in a large part of the delocalized phase. In both cases, one starts from a general exact formula (\ref{def}).  On the Cayley tree, the appropriate order of limits ($\eta\to 0$ at fixed large $N$) leads to the iterative procedure in terms of the recurrence relation (\ref{recsimple}), see the analysis  in Sec.~\ref{s2}. This procedure corresponds to the far asymptotic domain in terms of the recurrence relation (\ref{recurse}). In other words, the fixed point of Eq.~(\ref{recurse}) is not reached in view of the above order of limits and is thus immaterial for the statistics of eigenfunctions in the intermediate phase on the Cayley tree.  On the other hand, in the case of tree-like graphs without boundary of size $N>N_c(W)$, the statistics is controlled\cite{sparse} by the saddle-point of the corresponding supersymmetric action which is a solution of the self-consistency equation analogous to the fixed point of Eq.~(\ref{recurse}). Obviously, this argument applies also to many other observables, such as the level statistics or correlations of close-in-energy eigenfunctions.

On a more intuitive level, the importance of large-scale  loops (that distinguish the RRG and other tree-like models without boundary from the Cayley tree) for the eigenfunction statistics can be understood in the following way. To probe the statistics of a single eigenfunction on a RRG, we have to probe the physics on an energy scale of the order of level spacing $\delta_N$, i.e., on the time scale of the order of $N$. On the other hand, the typical size of the loops is $\ln N$. Therefore, it is expected that the loops may matter. 

\item
In the introductory part of the paper (Sec.~\ref{s1}), we pointed out that the interest to models of Anderson localization on tree-like structures  is at present largely motivated by their relation to problems of many-body localization. 
In view of the crucial difference between the Cayley tree and the RRG problems emphasized in the present work, it is 
natural to ask which type of model (Cayley tree or RRG) actually arises (at least within some approximations) when one characterizes the Fock-space structure of many-body eigenfunctions. We argue that RRG is more relevant in this context. Indeed, within the mapping of a many-body problem onto an effective tight-binding model, the vertices of an effective lattice represent basis many-body states of a free theory and the links represent interaction-induced matrix elements between them. Clearly, all typical basis states are equivalent, in the sense that each of them is connected to roughly the same number of other states. Thus, the effective model has no boundary and, in this sense, is analogous to RRG. It is worth emphasizing that, in a general situation of an extended system with spatially localized states, the effective model with be nevertheless essentially different from RRG, in view of spatial constraints on matrix elements. On the other hand, for problems of many-body localization in quantum-dot-type systems\cite{gornyi16,Gutman16}, a mapping to RRG may be a very good approximation; see, in particular, the corresponding arguments in Ref.~\onlinecite{Gutman16}.  Clearly, further studies of connections between RRG (and similar models on tree-like structures without boundary) with models of many-body localization is of great interest. It would be also interesting to see whether the Cayley tree problem might also find any application in this context.

\item
Very recently, a preprint appeared\cite{altshuler16}, the authors of which implement numerically an iterative procedure (within a certain population dynamics scheme) for a distribution of imaginary part of Green functions (i.e., of LDOS) on a Bethe lattice, with an idea to explore the limiting case of $\eta\to 0$ at fixed $N$. This procedure is thus analogous to the one that we implement analytically in Sec.~\ref{s2} to study wave function statistics at a root of a finite Cayley tree with a boundary. The authors of Ref.~\onlinecite{altshuler16} do find in this way a fractal behavior, in consistency with our results. However, they argue that what is studied in this way is the eigenfunction statistics on RRG rather than on a Cayley tree. As we have shown, this interpretation is erroneous. 

\item
The present work opens a way for exploring further statistical properties of the peculiar fractal delocalized phase on a Cayley tree. Let us give some examples. 
First, one can study the eigenfunction statistics away from the root. We expect  \cite{to-be-published} that the fractal exponents depend on the position of the observation point, in a certain analogy with the boundary multifractality at Anderson transitions\cite{evers08}.  In fact, it was found in Ref.~\onlinecite{aizenman06} that random Schr\"odinger operatirs on certain ``canopy graphs'' have pure-point spectrum for any strength of disorder, at least for some models of disorder distribution. This suggest localization of eigenstates near the boundary of a Cayley tree. It would be very interesting to see how this localization manifests itself in the eigenfunction statistics near the boundary and how this behavior crosses over to the delocalization (ergodic or fractal) near the root of the tree. Our preliminary results \cite{to-be-published} indeed  indicate that even in the weak-disorder phase, $W<W_e$ (or $g > g_e$ in the sigma-model language), each eigenstate on the is localized near a single path connecting a root with the boundary. Thus, even the phase $W<W_e$ ($g > g_e$) on the Cayley tree is ``ergodic'' only from the point of view of eigenfunction statistics at the root.
Second, it would be interesting to study wave function correlations (in space and in energy) that are expected to be unusual (see a discussion in Sec.~\ref{s2c}).

\item Similarly to the problem of Anderson localization, various formulations of spin-glass theory on tree-like graphs were discussed\cite{kitaev91,mezard01,lucibello14}. This includes, first, a finite Cayley tree with boundary and with $L$ generation such that correlations are studied only around the root of the tree (within a distance $L'$ from the root, with $L/L' \to\infty$). Second, models on tree-like structures without boundary  but with large loops (RRG and SRM) were studied. It was found the latter formulation (RRG or SRM) is advantageous; in this case the problem is described by the self-consistency equation and the finite-size correction can be analyzed\cite{mezard01,lucibello14}. The problems of Anderson localization and of spin glass have common property of being characterized by an order-parameter function, and the analysis of the spin glass on RRG in Ref.~\onlinecite{lucibello14} bears certain similarity with the theory of Anderson localization on SRM in Ref.~\onlinecite{sparse}. As to the model of a spin glass on a Cayley tree with a boundary, it was found\cite{mezard01} that it suffers from some ambiguity since the dependence on boundary conditions remains, even in the limit $L/L'\to\infty$, because of frustration. Thus, there appears to be a certain similarity between our finding of a qualitatively different behavior of the eigenfunction statistics at a root of a Cayley tree, on one hand, and in RRG, on the other hand, and an analogous difference in spin-glass models. It remains to be seen whether this analogy has deeper physical roots.

\end{enumerate}

\section{Acknowledgment}

We thank  M.V. Feigelman, L.B. Ioffe, V.E. Kravtsov, and A. Kitaev for interesting discussions during the Chernogolovka conference in June 2016 where this work was presented. We also thank Y.V. Fyodorov for useful comments.
The work was supported by the Russian Science Foundation under Grant No.\ 14-42-00044 
and by the EU Network FP7-PEOPLE-2013-IRSES under Grant No. 612624 ``InterNoM''.


\begin{thebibliography}{99}
\bibitem{note-beta} For $W=14$ we used the numerically extracted value $\beta_* \simeq 0.72$; for $W=8$ and $W=5$ we used the estimates $\beta_*\simeq 0.85$ and $\beta_* \simeq 1.0$, respectively.
\end{thebibliography}
\end{document}